\documentclass[11pt,onecolumn]{article}

\usepackage{amsfonts}
\usepackage{amsmath}
\usepackage{amssymb}
\usepackage{bm}
\usepackage{graphicx}
\usepackage{wrapfig}
\usepackage[font=small,labelfont=bf]{caption}
\usepackage{xspace}
\usepackage{algorithm}
\usepackage{algorithmic}
\usepackage{color}
\usepackage{subcaption}

\long\def\remove#1{}

\newcommand{\etal}      {et al.\@\xspace}
\newcommand{\geo}		{\mathsf{g}}
\newcommand{\KK}		{\mathsf{K}}		

\newcommand{\tile}		{{\Omega}}
\newcommand{\stable}		{\mathsf{G}} 

\newcommand{\R}		{\mathbb{R}}	

\definecolor{darkred}{rgb}{1, 0.1, 0.3}
\definecolor{darkblue}{rgb}{0.1, 0.1, 1}
\definecolor{darkgreen}{RGB}{0, 102, 0}

\newcommand{\myimage}	{{I}} 
\newcommand{\Ghat}		{{\widehat{G}}}
\newcommand{\Vhat}			{{\widehat{V}}}
\newcommand{\Ehat}			{{\widehat{E}}}
\newcommand{\myroot}		{{\mathrm{v}}}
\newcommand{\disV}			{{\mathrm{W}}}

\begin{document}
	
\title{Topological Skeletonization and Tree-Summarization of Neurons Using Discrete Morse Theory}
\author{Suyi Wang\thanks{Computer Science and Engineering Department, The Ohio State University, Columbus, OH 43210.} \and Xu Li\thanks{Cold Spring Harbor Laboratory, Cold Spring Harbor, NY 11724} \and Partha Mitra$^\dagger$ \and Yusu Wang$^*$}
\date{}
\maketitle

  \begin{abstract}
  
Neuroscientific data analysis has classically involved methods for statistical signal and image processing, drawing on linear algebra and stochastic process theory. However, digitized neuroanatomical data sets containing labelled neurons, either individually or in groups labelled by tracer injections, do not fully fit into this classical framework. The tree-like shapes of neurons cannot mathematically be adequately described as points in a vector space (eg, the subtraction of two neuronal shapes is not a meaningful operation). There is therefore a need for new approaches. Methods from computational topology and geometry are naturally suited to the analysis of neuronal shapes. Here we introduce methods from Discrete Morse Theory to extract tree-skeletons of individual neurons from volumetric brain image data, or to summarize collections of neurons labelled by localized anterograde tracer injections. Since individual neurons are topologically trees, it is sensible to summarize the collection of neurons labelled by a localized anterograde tracer injection using a consensus tree-shape. This consensus tree provides a richer information summary than the regional or voxel-based "connectivity matrix" approach that has previously been used in the literature. 

The algorithmic procedure includes an initial pre-processing step to extract a density field from the raw volumetric image data, followed by initial skeleton extraction from the density field using a discrete version of a 1-(un)stable manifold of the density field. Heuristically, if the density field is regarded as a mountainous landscape, then the 1-(un)stable manifold follows the "mountain ridges" connecting the maxima of the density field. We then simplify this skeleton-graph into a tree using a shortest-path approach and methods derived from persistent homology. The advantage of this approach is that it uses global information about the density field and is therefore robust to local fluctuations and non-uniformly distributed input signals. To be able to handle large data sets, we use a divide-and-conquer approach. The resulting software {\it DiMorSC} is available on Github\cite{DiMorSC}. To the best of our knowledge this is currently the only publicly available code for the extraction of the 1-unstable manifold from an arbitrary simplicial complex using the Discrete Morse approach.

  \end{abstract}
  
  \section{Introduction}
	Understanding the neuronal connectivity architecture of brains is an important goal in neuroscience. The primary approach brought to bear on this goal is the reconstruction of neuronal projections following injections of tracers within the brain. With the development of high throughput pipelines and technologies that can deal with large digital data sets, it is now possible to analyze whole brain datasets at a cellular resolution. However, there is a need for developing new methods that can facilitate the modeling and understanding of neuronal morphology and thus aid in extracting connectivity information from brain image volumes. The typical approach is to summarize the results in the form of regional, or voxel to voxel "connectivity matrices" or directed graphs, with the source region given by the tracer injection site and the target region containing tracer labelled axons or retrogradely labelled neurons. However, this connectivity matrix summary loses all information about the tree-like structure and shape of the projection neurons that constitute the tracer-labelled set. Here we introduce a conceptually distinct summary of an anterograde tracer injection (which labels a collection of neuronal somata concentrated at the injection site, together with the projecting axons and dendrites) in the form of a consensus tree that provides a geometrically and topologically meaningful summary of the collection of neurons labelled by the tracer injection.

	%
		The methodology developed here is general and applies also to the idealized case of a single labelled neuron, and therefore provides an additional method for skeletonization individually labelled neurons from volumetric image data sets. A large number methods exist for automatic neuron tracing, mostly for reconstructing single neurons, 
	such as \cite{Al-Kofahi:2002, Bas2011, Basu2014, Boykov2001, Choromanska2012, Chothani2011, activelearning2014, Lee2008, lee2012, Myatt2012, oh2014mesoscale, sironi2015, SCHMITT20041283, sui2014, Srinivasan2007, turetken2013, Turetken2011, VASILKOSKI2009197, Yang2013, ZHANG2007149, Zhao2011, Zhou1999, Zhou2015, Zhou2016}; see also surveys \cite{Acciai2016, DONOHUE201194, neuronsurvey} and book \cite{arenkiel2014neural} for more comprehensive discussions on neuron tracing methods. The advantage of the Discrete Morse based skeletonization for single neurons arises from the usage of global (as opposed to purely local) information, leading to reconstructions that have robustness to local variations and non-uniform signal distributions. 

\begin{figure}
		\centering
		\includegraphics[width=0.9\textwidth]{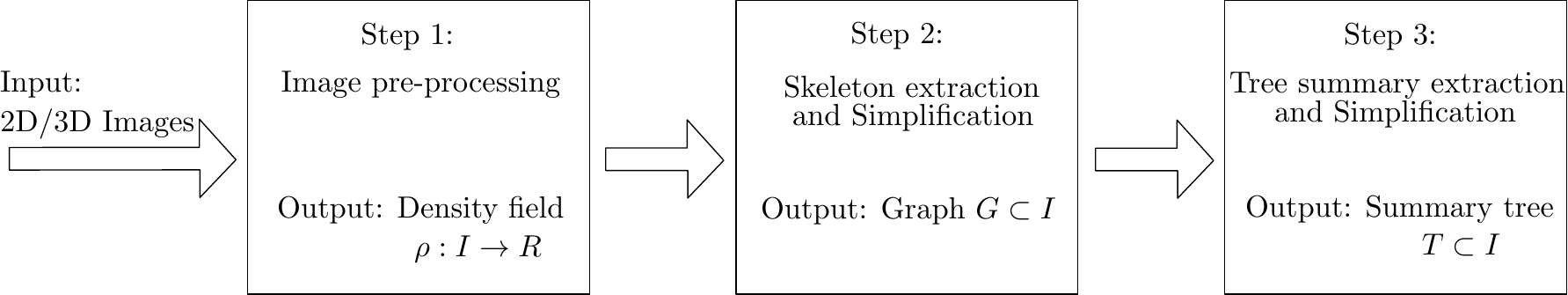}
		\caption{Computational fipeline for our tree-summarization framework of an input image.} 
%
		\label{fig:pipe}
	\end{figure}

\paragraph{Current work.}	In this paper, we propose a pipeline to summarize the 3D image stacks  based on topological methods. We apply the pipeline to trace neuronal projections following the anterograde injection of AAV tracer.  We also provide the details about the accompanying software, DiMorSC. Our pipeline has three stages (see Figure \ref{fig:pipe}). It first performs a pre-processing  step and converts the input 3D image stack into a density field. 
It then extracts the skeleton from the density field using the topological concept of \emph{1-(un)stable manifolds} from discrete Morse theory.
Previously, 1-stable manifolds have been successfully applied to extract graph skeletons from 2D/3D data, such as extracting cosmic web from
	the simulated density of dark matter in $\R^3$~\cite{2011MNRAS} and the reconstruction of road networks from a large collection of GPS trajectories \cite{Wang2015}. 
We adapt this idea for summarizing 3D mesoscopic AAV tracer images. 
In order to handle the relative large size of image data, we propose a scalable divide-and-conquer strategy to compute this skeleton.
     Finally, after an initial graph skeleton is extracted, we develop a shortest path based approach to convert the skeleton to a summary tree which is rooted at the injection site. 
     We further provide a simplification strategy, via ideas from the persistent homology from computational topology, to control the level of details of the final summarization.  
	Our proposed method works directly on 3D image volumes. 
	Leveraging the topological structure behind the density field, our method uses the global information from the input images and thus can model and summarize global connectivity paths resulting from tracer injections. It is robust to noise as well as non-uniformly distributed input signals. 
	
The implementation of the algorithm presented here, is based on discrete Morse theory, which extracts the skeleton in a \emph{combinatorial manner} rather than in a numerical manner. 
Overall, our approach provides a unified framework with a mathematical foundation (based on discrete Morse theory combined with topological persistence) for  extracting a tree skeleton from input images. The \emph{combinatorial nature} of our algorithm, combined with the \emph{systematic simplification strategy}, allow us to extract a principal backbone from noisy input images containing complex signals. 
We compare results obtained from the topological approach to existing methods for skeletonization single neurons and show that it provides comparable or better performance (Section \ref{subsec:exp:single}). 
We apply these methods to the novel summarization of anterograde tracer injection data from the Mouse Brain Architecture Project to map whole-brain connectivity at a mesoscopic scale. 
Compared with previous regional connectivity based data summaries, the tree-summary retains geometrical and topological information about the projection patterns lost in the connectivity matrix summary. This should be also useful in connecting the tracer-injection data with single neuron projection data from the injection sites, as the tree summary provides a consensus tree structure that captures information about the population of neurons with somata localized at the injection site. 
     The resulting software is publicly available at \cite{DiMorSC}.
	
The remainder of the paper is organized as follows:	
Below we first briefly discuss some related work. 
	In Section \ref{sec:method} we introduce our pipeline for neuron tracing and describe details of our method.
	In Section \ref{sec:stitching}, we discuss the divide-and-conquer strategy to handle large data set that cannot fit in the available memory.
	Finally in section \ref{sec:result}, we first demonstrate the effectiveness of our proposed software on single neuron reconstruction that we tested on datasets freely available from the DIADEM challenge \cite{diademchallenge}. We then show results on summarization of anterograde tracer injetions. 


\paragraph{Related work.}
There is a large literature for single neuron reconstruction e.g \cite{Al-Kofahi:2002, Bas2011, Basu2014, Boykov2001, Choromanska2012, Chothani2011, activelearning2014, Lee2008, lee2012, Myatt2012, oh2014mesoscale, sironi2015, SCHMITT20041283, sui2014, Srinivasan2007, turetken2013, Turetken2011, VASILKOSKI2009197, Yang2013, ZHANG2007149, Zhao2011, Zhou1999, Zhou2015, Zhou2016}. 
We refer the readers to surveys \cite{Acciai2016, DONOHUE201194, neuronsurvey} and a book \cite{arenkiel2014neural} for more comprehensive discussions on (single) neuron tracing methods, and we only mention a few representative ones below. We note that many of the algorithms are incorporated into the publicly available visualization platform Vaa3D \cite{Vaa3D}. 

Most of the neuron tracing algorithms either grow a neuron sequentially 
	(such as growing a tree from the root), or connect (certain special) points in a non-sequential manner. 
    A popular class of sequential tracing algorithms uses a shortest-path based approach, 
	such as APP\cite{pengapp}, APP2 \cite{app2} and SmartTracing \cite{smarttracing}.
    These methods start with a given seed point and grow it into a tree by 
    connecting new nodes to the existing tree through the shortest paths. 
	For example, in APP2, all pixels with intensity lower than a predefined threshold are treated 
	as background and the foreground pixels are processed by a fast marching method, 
	which computes for every pixel its shortest distance to the background.
	The distance field of foreground is then used as the new 3D image, and intuitively, the intensity of pixels lying in the middle of a neuron cell is higher than those close to the boundary. 
	After fast marching, APP2 sets a base point and computes the 
	shortest path tree to the base point as an initially reconstructed neuron tree.
	The tree is then pruned according to the lengths of the branches.
	In SmartTracing \cite{smarttracing}, 
	the result is further improved by a machine learning approach that better classifies 
	a pixel as background, foreground or unknown.
	APP2 is efficient, however, the algorithm could stop at a gap in the signal.
	The authors resolve this issue by developing a search procedure to explore around the gap.
	The machine learning based SmartTracing potentially generates more accurate results, but is significantly slower than APP2. For our comparisons, we assume that APP2 and SmartTracing represent the state-of-the-art algorithms for single neuron tracing.

	The tracing can also be performed sequentially only at the branch level. For example, in the  active contour based approach \cite{Wang2011} proposed by Wang et. al., 
a gradient vector is computed for each pixel in 3D.
	The authors then define an energy function using the gradient vector as the external force and a smoothness function as the internal force.
	The method iteratively identifies a foreground point using
	Frangi’s vesselness measure \cite{Frangi1998} and grows it to an arc on the neuron tree
	while minimizing the energy function.
	Upon obtaining a maximal arc, all nearby points are removed and 
	this growing procedure is repeated until all foreground points are traced.
    Compared to the global sequential tracing methods such as APP2, this type of ``semi-sequential tracing'' (at the branch level) may be more accurate in generating individual branches, but assembling the branches into a single tree can be challenging.
	
    
    The above approaches usually require the segmentation of the foreground from the background within the image. The quality of the reconstruction is dependent on the quality of this segmentation. For images that are noisy, i.e., where intensity of signals are not always homogeneous, the segmentation quality often is problematic and tends to cause gaps and  can even lead to broken branches.  
    
    As described in {\bf Our work}, our approach resolves this issue by taking a global view of the entire data, and traces the neuron based on the topological structure behind the given input 3D image data. 
Our neuron tracing approach builds upon the algorithm developed by Yuan \etal \cite{Yuan2009},
	which traces neurons by connecting the critical points in the input. 
	However, there are major differences.
	(1) In Yuan {\it et al}, the critical points are connected by the trace induced by 
	moving the saddle points along the gradient using a numerical method, 
	while our approach uses the topological concept of ``1-stable manifold'' 
	from Morse theory to connect them in a theoretically well-founded manner.
	Furthermore, we use the \emph{discrete Morse theory} to compute such 1-stable manifolds in 
	a robust combinatorial manner (in contrast to a numerical approach).
	(2) Yuan {]it et al} simplifies the initial result using a shortest path tree and 
	further prunes the tree by an erosion scheme, 
	while our approach uses topological persistence simplification in a systematic manner, and less important features are removed first.
	In summary, our approach provides a unified, conceptually simple, and mathematically sound framework to extract a tree skeleton from input image volumes. The combinatorial nature of our algorithm, combined with the systematic simplification strategy, allows us to extract the main backbone from noisy input images with complex signal content. 

	From a computer science perspective, our apporach builds upon prior work on using discrete Morse based methodology for extracting skeletons of images; see e.g, \cite{Gyulassythesis,DRS15,RWS11,2011MNRAS,Wang2015}. Our software is based on the work of Gyulassy \cite{Gyulassythesis, GDN07}, of Sousbie \cite{2011MNRAS}, and Wang et al. \cite{Wang2015}. Our algorithm focuses on the specific case of extracting graph skeletons from input data, and simplifies the previous approaches for this specific case (e.g., we do not need to handle the cancellation of edge-triangle pairs during the simplification stage). Our implementation can also take an arbitrary simplifical complex as input, while the previous implementation works for 2D / 3D images or Delaunay triangulations (which is a specific type of simplicial complex). This apporach of using an arbitrary simplicial complex input,  improves the efficiency in handling large images, as the arbitrary simplicial complex allows us to consider only regions around signal, which can be rather sparse within the input images. 
    Finally, this graph skeleton reconstruction step is combined with a tree-extraction and simplification step to produce the final summary.

\section{Method}
	\label{sec:method}

	\subsection{Overview}
\begin{figure}
			\centering
            \begin{tabular}{ccccccc}
            \includegraphics[width=0.2\textwidth, height = 4cm]{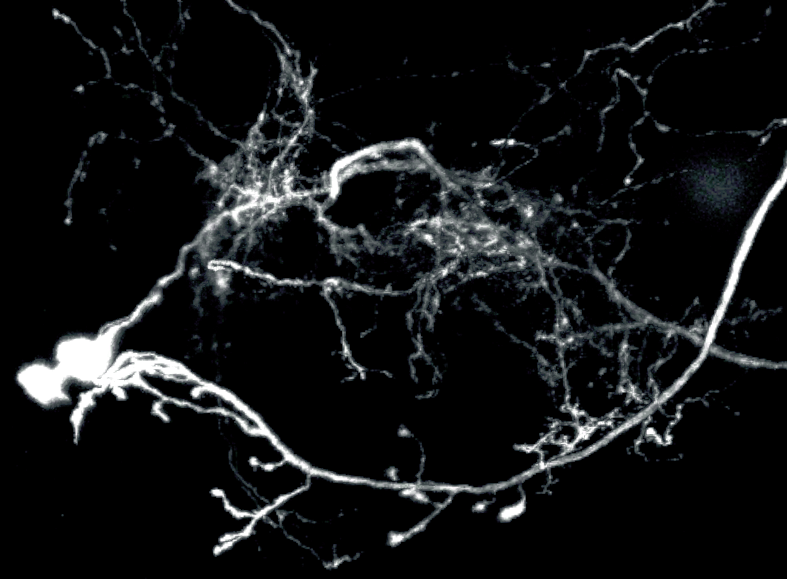} & \hspace*{0.0in} &
            \includegraphics[width=0.2\textwidth, height = 4cm]{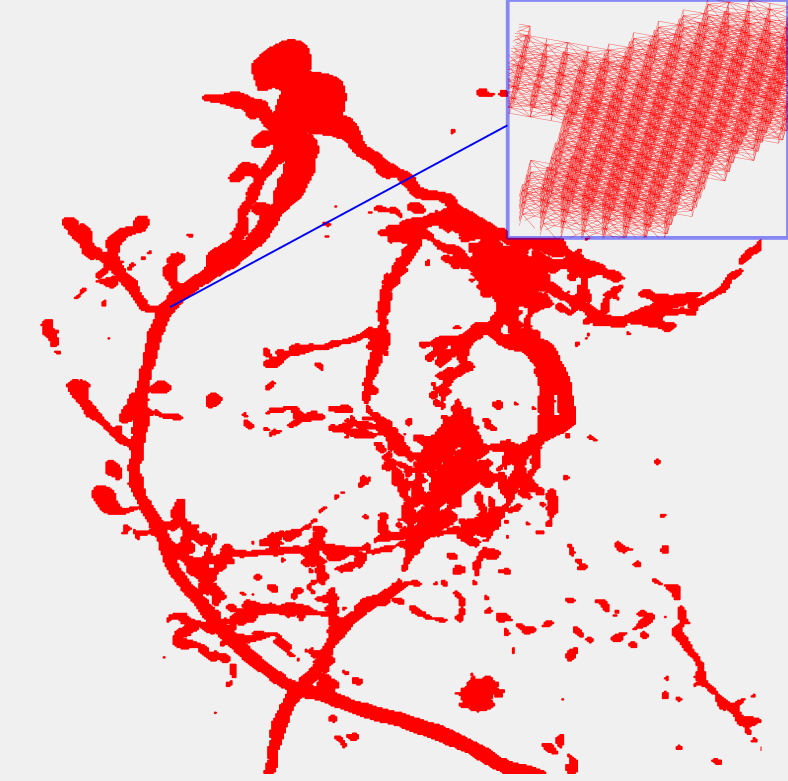} & \hspace*{0.0in} &
            \includegraphics[width=0.2\textwidth, height = 4cm]{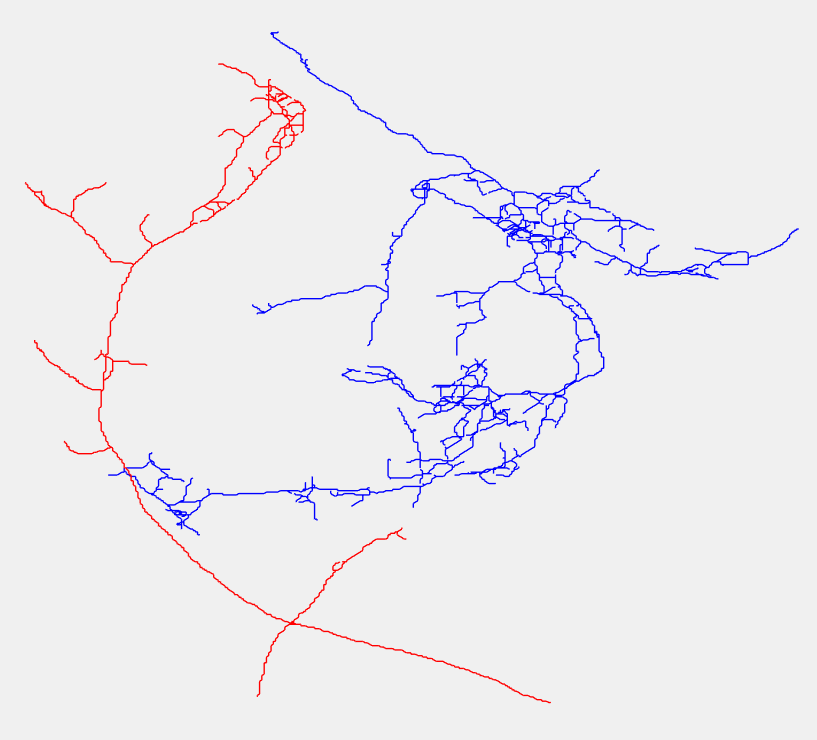} & \hspace*{0.0in} &
             \includegraphics[width=0.2\textwidth, height = 4cm]{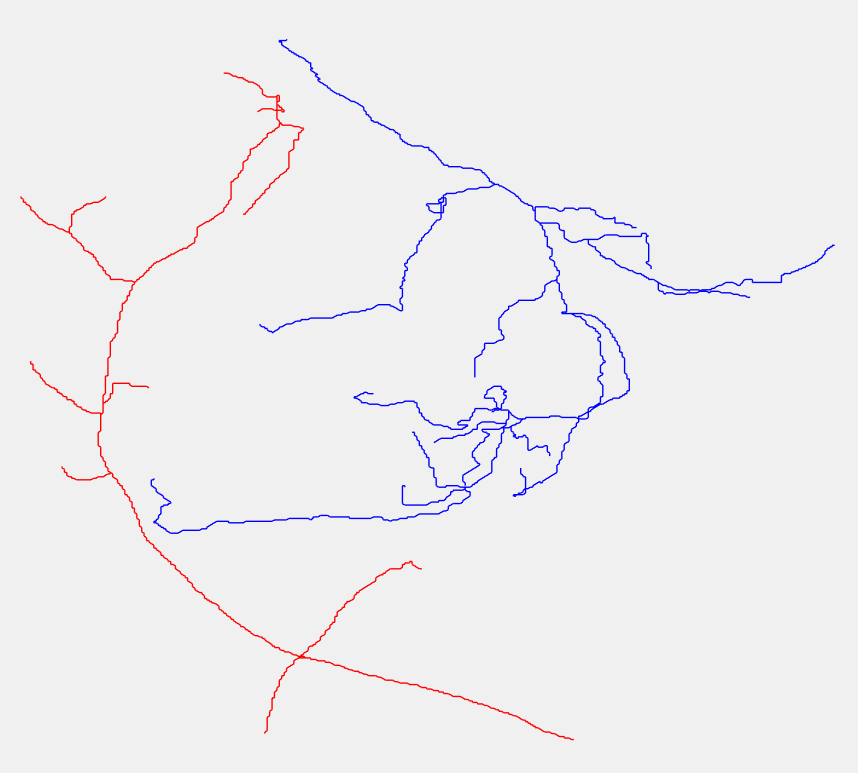} \\
             (a) & & (b) & & (c) & & (d)
             \end{tabular}
			
			  \caption{Pipeline overview: the Input image volume in (a) is first converted to the 3D density map shown in (b).  
              (c): initial graph extraction, and (d): tree extraction and simplification 
			  (red and blue trees are the final two output trees, reconstructing two neurons); all in 3D. 
			 \label{fig:pipe_detail}}
		\end{figure}

		Our pipeline accepts 3D images and outputs a geometric tree in three steps as shown in Figure \ref{fig:pipe}:
		image processing to convert the input image to a density field $f: \myimage \to \mathbb{R}$ defined on the 3D cube $\myimage$; extracting graph skeleton $G$  from $f$, and tree-summary $T$ extraction and simplification from $G$. 
        An example illustrating the pipeline is given in Figure \ref{fig:pipe_detail}.

		\paragraph{{\sf Step 1: }Image pre-processing.} This step converts the input from 3D image stacks to a density (grey scale) map. 
The density map $\rho: \myimage \to \mathbb{R}$ is a function defined on the 3D cube $\myimage = [0,1]^3$. In the discrete case, this domain $\myimage$ is represented by a cubic grid, which is further triangulated and represented by a so-called \emph{simplicial complex} $\KK$, consisting of a set of vertices, edges, triangles and tetrahedra. 
However, as we will see later: (i) we only need the vertices, edges and triangles of $\KK$ for our graph skeleton and tree-summary extraction; thus from now on, we assume that $\KK$ consists of only the vertices, edges and triangles from the triangulation of $\myimage$. (ii) In cases when the input image is large in size, we can restrict $\KK$ to a sub-complex of it which intuitively captures where signals lie. 
		Given a triangulation $\KK$ of $\myimage$, the density map $\rho$ is defined at vertices of $\KK$, and in what follows, we sometimes refer to it as the density map $\rho: \KK \to \mathbb{R}$. 
	Details of this step are described in Section \ref{subsec:preprocessing}. 


		\remove{
		\paragraph{Triangulating.} This step creates a triangulation for 
		pixels with intensity higher than a given threshold. 
		Note that the threshold here should be low,
		removing only obvious background to reduce the size of data.
		This step is not intended for signal extraction because signal is non-homogeneous.
		Since the image data from the input is on a cubic grid, when triangulating, 
		we only connect the data points to their immediate neighbourhood.
		Figure \ref{fig:pipe_detail}B shows the output of the triangulation step.
		Note that the figure shows only the triangulation and the actual function values are not visualized.
		}

		\paragraph{{\sf Step 2: }Graph skeletonization.} This step extracts the graph skeleton of the density map $\rho: \KK \to \mathbb{R}$ using the so-called 1-stable manifolds of $f$, which are computed via the discrete Morse theory. 
		Intuitively, if we view the graph of the density map as a terrain defined on $\mathbb{R}^3 \times \mathbb{R}$, the 1-stable manifolds capture the network of ``mountain ridges'' of this terrain, representing the ``center curves'' of the local high density regions (corresponding to signals). 
		This concept and its computation will be described in Section \ref{subsec:skeletonization}.
		
		\paragraph{{\sf Step 3: }Tree extraction and summarization.} 
		Given the graph skeleton $G$ extracted in {Step 2}, this step converts it to a tree summary $T$: This tree can be rooted at a specific choice of root if desired (say the injection site in the input AAV tracer image). We further develop a simplification strategy to simply this tree and control its level of details in a systematic manner, using ideas from topological persistence. 
Details will be presented in Section \ref{subsec:treesimp}.

	\subsection{Step 1: Preprocessing} \label{subsec:preprocessing}
		
        
        The input is a 3D image volume $\myimage$ consisting of a stack of 2D images. The purpose of this step is to extract a density map from $\rho: \myimage \to \mathbb{R}$, where the value of $\myimage$ is given at discrete grid points (pixels of the input images) and indicates the strength of the signal at this point. 
        
        The input image can be of different types. Potentially different image pre-processing techniques will be needed to handle different type of input images.   
  		The procedure we describe here was developed for fluorescent image stacks of mouse brains injected with a tracer substance (Adeno-Associated Viruses carrying a Green or Red Fluorescent Protein payload) scanned on a Whole Slide Imager (a Hamamatsu Nanozoomer). The raw data consists of a 3D stack of RGB images, 12 bits/color channel, in-plane resolution of $0.46\mu$ and section spacing of $40\mu$. The preprocessing for quantifying the green and red tracer are performed separately. For the green tracer, a Laplacian of Gaussian(LoG) filter is applied first to increase the signal to background ratio and sharpen the tracer signal. Then the image is converted to HSV and LAB colorspace to filter out noisy background. 
Blood vessel artifacts are detected and removed using Circle Hough transform. The images are registered onto a reference atlas using procedures described elsewhere. 
	
\begin{figure}
	\centering
	 \includegraphics[width=0.9\textwidth, height=10cm]{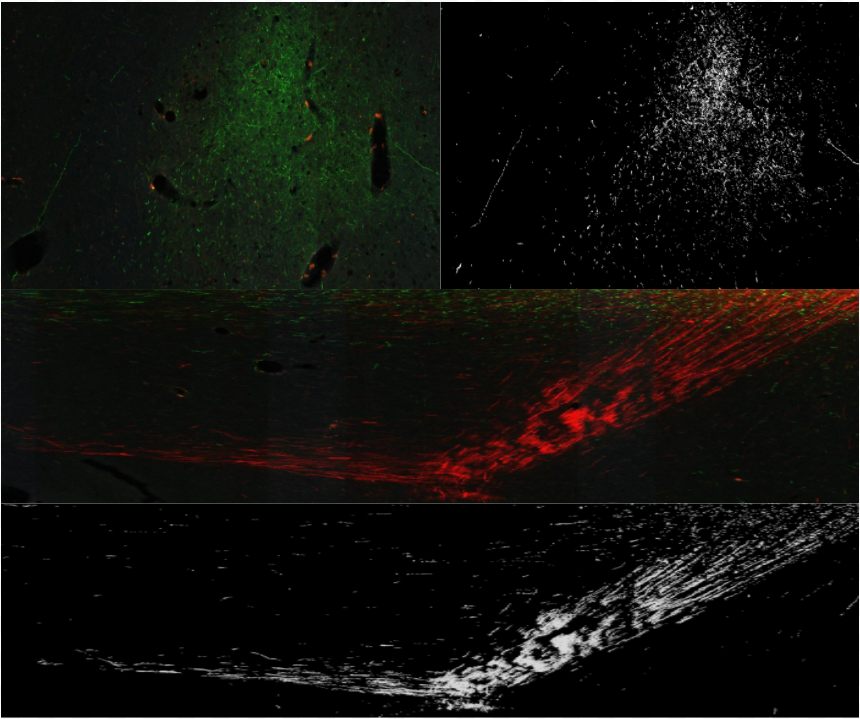}
	\caption{Preprocessing results for green (top two images) and red AAV tracer (bottom two images). The green and red signals result from fluorescent tracers visible in part of a tracer injected mouse brain. The black and white images are resulting density maps. }
	\label{fig:PMD_pre}
	\end{figure}

    For the red tracer, due to the presence of autofluorescence that is not related to the signal\cite{autofluoro} in the brain, a simple k-nn clustering strategy was used to  separate tracer signal from autofluorescence noise. In addition, the autofluorescence can be regarded as shot noise in direction perpendicular to the sections, so a median filter is also used. 
An example of the preprocessing result is demonstrated in Figure \ref{fig:PMD_pre}. We refer to the initial density field after this pre-processing as $\rho_1: \myimage \to \mathbb{R}$. 
    
    For later steps, we assume that our input domain is modeled by a triangulation $\KK$ (consisting of vertices, edges, and triangles \footnote{Strictly speaking, the triangulation of the 3D domain also include tetradedral cells. However, for our later algorithm, only the so-called 2-skeleton, which consists of vertices, edges and triangles, is needed and thus computed.}) instead of having cubic cells. 
    We consider the pixels as points on a regular grid and triangulate each cubic cell (the exact triangulation is not important, as long as triangulating neighboring cubes gives rise to consistent triangulation of the common (square) face they share). 
		
	Now we have a triangulation $\KK$ of $\myimage$ with an initial density function $\rho_1: vert(\KK) \to \mathbb{R}$ defined at vertices $vert(\KK)$ of $\KK$ (which are the grid points in $I$). By default, our algorithm performs one more smoothing step, by smoothing $\rho_1$ with a Gaussian kernel within a small neighborhood of each point. Let $\rho: vert(\KK) \to \mathbb{R}$ be the resulting final density map. 

	We perform this smoothing stage for the following reason:
    	The initial density map $\rho_1$ might contain a signal plateau (flat top) area, 
		on which the 1-stable manifold (mountain ridge) in the next step could be ambiguous. 
		For example, in places where the signal is saturated, there could be a thick band of pixels  
		with the same (highest) value, form a plateau (flat mountain peak) in the terrain formed by this function. 
This gives a degenerate case for defining/tracing ``mountain ridges'' in our later steps. 
        The Gaussian smoothing help alleviate the situation: it would strengthen the signal in the center of such flat regions, while reduce the intensity of pixels in the boundary region. 
        Furthermore, the preprocessing strategy may segment the input image and output a binary density field where points in the foreground have value 1 and those in the background having value 0. The Gaussian smoothing converts such an input into a smoother field, with points along ``centerlines'' of the foreground having higher function values, so that the ``mountain ridges'' of such a terrain can later be captured by our 1-stable manifolds approach. 
        

		Finally, we remark that in our pipeline, instead of using the triangulation $K$ of the entire 3D image $I$, we can also use a subcomplex $\KK' \subset \KK$ spanned by only pixels whose density value is larger than a threshold to reduce the size of input. Note that if needed, a very low threshold can be used to remove the points that are obviously background, so as not to cause gaps in the remaining subcomplex. 
Allowing for adequate computing resources, this method would be more reliable on the full image without the removal of any data point. 

	\subsection{Step 2: Graph skeletonization}
		\label{subsec:skeletonization}

		The input to this step is a density field $\rho: \KK \to \mathbb{R}$ where $\KK$ is a triangulation of $\myimage \subset \mathbb{R}^3$. Below we first explain the main idea for the continuous setting where $\rho$ is assumed to be a smooth function defined on the domain (3D cube) $\rho : \myimage \to \mathbb{R}$.



		Formally, given any smooth function $f:\R^3 \rightarrow \R$, the gradient of a point $p\in \mathbb{R}^3$, $\nabla f(p) = -[\frac{\partial f}{\partial x}, \frac{\partial f}{\partial y}, \frac{\partial f}{\partial z}]^T$ indicates the direction at $p$ along which the rate of change of the function $f$ is largest. A point $p \in \R^3$ is \textit{critical} if the gradient at $p$ is the zero vector, otherwise $p$ is \textit{regular}. 
	In Morse theory \cite{Morse_Theory}, if the input function $\rho$ is sufficient nice (more formally, it is a Morse function, ie with non-zero Hessians at the critical points), then there are four types of non-degenerate critical points for $\rho$ defined on $\mathbb{R}^3$ - maxima, minima and two types of saddles of index 1 and 2, respectively. 
        %
%
%
%
%
\remove{
		\begin{figure}[hbtp]
		  \begin{center}
			\includegraphics[width=0.7\textwidth]{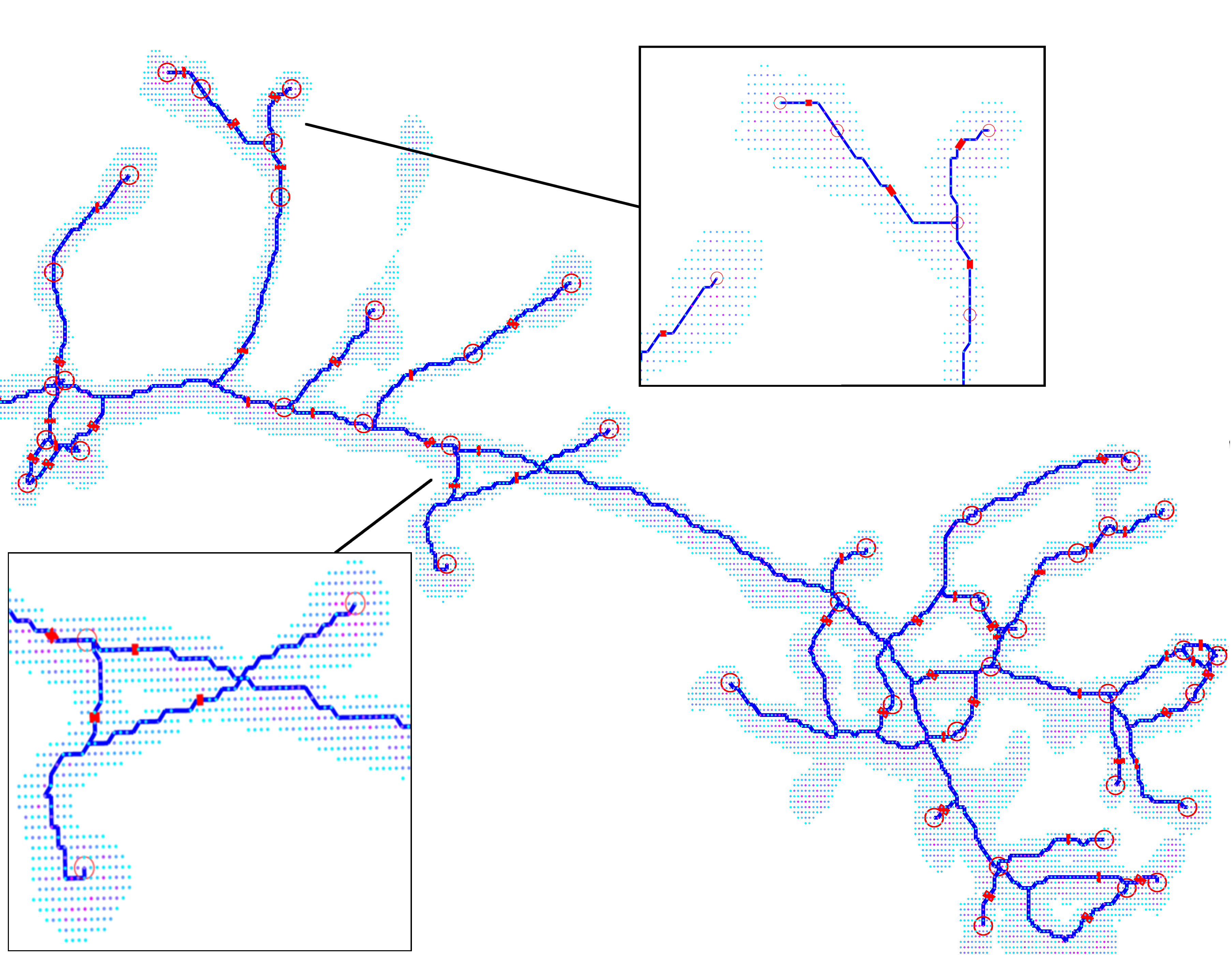}
		  \end{center}
		  \caption{Critical points as indicator of neurons.}
		  \label{fig:critical_to_neuron}
		\end{figure}
}
\begin{figure}[hbtp]
		  \begin{center}
\begin{tabular}{ccc}
			\includegraphics[width=0.5\textwidth]{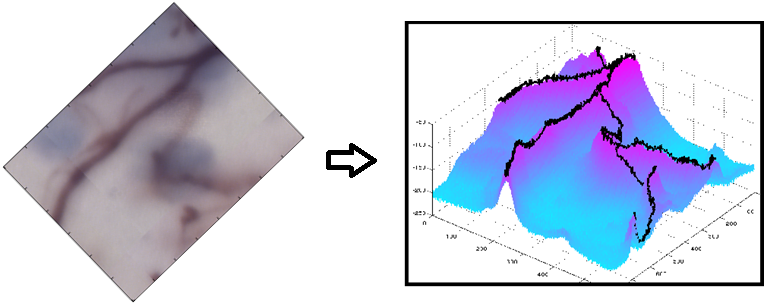} & &
	\includegraphics[height=3cm,width=0.35\textwidth]{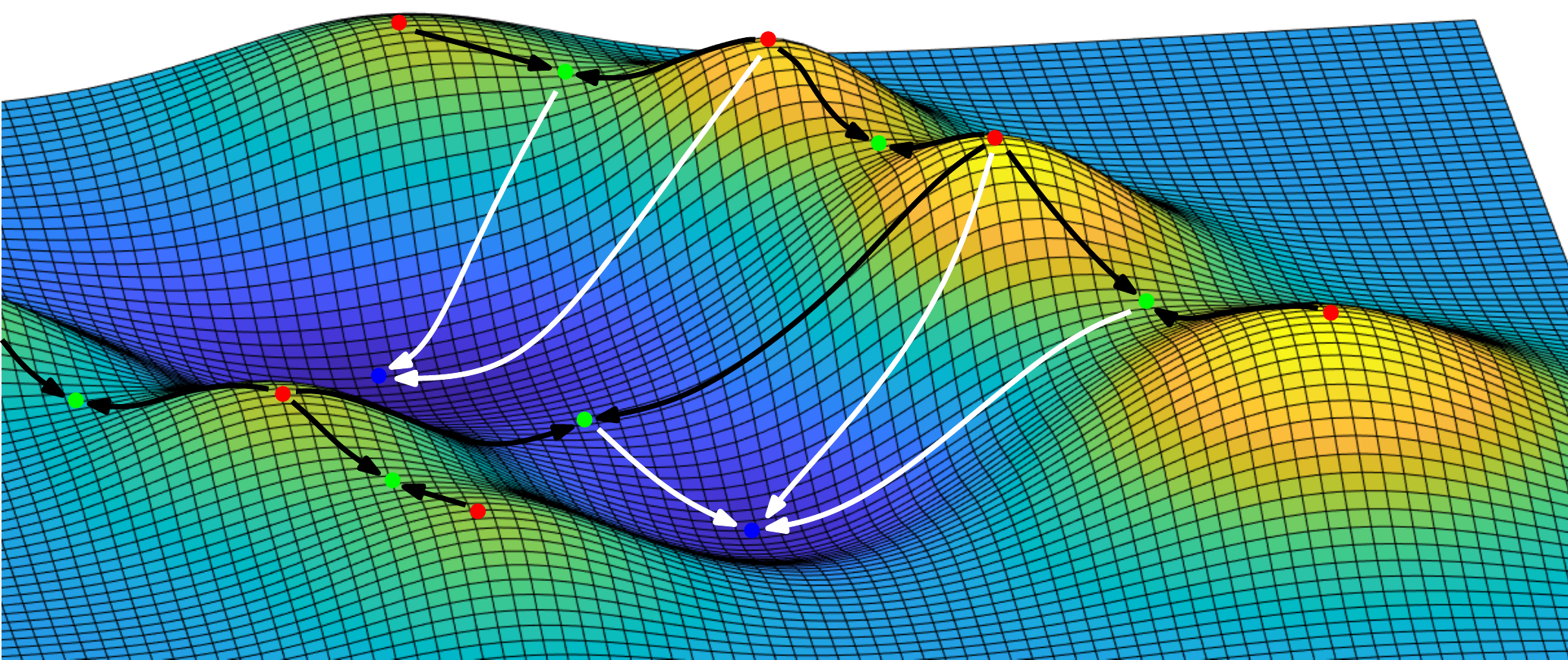} \\
(a) & & (b)
\end{tabular}
		  \end{center}
		  \caption{(a) An example of a 2D image converted to a density function with the graph (terrain) of this function shown in the right. (b) An example of 2D terrain (graph of 2D function): red points are maxima, green points are saddles, while blue points are minima. The white paths are some example of integral paths ending in minima. The black curves are a collection of 1-stable manifolds (between maxima and saddles).}
		  \label{fig:yusutmp}
		\end{figure}

	Intuitively, if we view the graph of the density function $f: \mathbb{R}^3 \to \mathbb{R}$ as a terrain defined on $\mathbb{R}^3 \times \mathbb{R}$, then maxima are the mountain peaks, while minima indicate valley basins. See Figure \ref{fig:yusutmp} where for illustration purpose, we provide an example for a 2D density function $f: \mathbb{R}^2 \to \mathbb{R}$: In this case, the terrain (graph of this function) is defined on $\mathbb{R}^2 \times \mathbb{R}$, consisting of points $(x, y, f(x,y))$ (that is, the height of a point $(x,y) \in \mathbb{R}^2$ represents its function value $f(x,y)$). For a function defined on $\mathbb{R}^2$, there are three types of non-degenerate critical points: maxima, minima and saddle points; see Figure \ref{fig:yusutmp} (b), where red dots are maxima, blue ones are minima, while the green dots are saddle points. 
We will use certain curves connecting maxima and index-2 saddle points in 3D case (or connecting maxima and saddle points for the 2D case) as a way to capture the hidden graph skeleton of this density function $f$. Below, we will first introduce the notations, and then provide the intuition behind the procedure. 

		An integral line $L:(0,1)\rightarrow \R^d$ is a maximal path in $\R^d$, where tangent vectors are consistent with the gradient for all points on the line.
		Imagine putting a drop of water on the terrain: it will flow downwards following the gradient direction at any moment; if we negate the function value, then it will move upwards following the negation of the gradient direction. The trajectory of it (both downwards and upwards) forms the integral line passing through that point. 
		The destination of an integral line is $dest(L) = \lim_{p \rightarrow 1}L(p)$ and the origin of an integral line is $ori(L) = \lim_{p \rightarrow 0}L(p)$. We also set by $dest(x)$ (resp. $ori(x)$) to be the destination (resp. the origin) of the integral line passing through $x$.
		The origin and destination of an integral line are necessarily critical points. See Figure \ref{fig:yusutmp} (b) for a 2D illustration. 
		The stable manifold of a critical point $p$ is defined as:
		$$S(p) = \{p\}\cup \{x \in \R^d \mid ~dest(x) = p\}.$$


		In other words, the stable manifold of a critical point $p$ is the union of itself and all points whose integral lines eventually flow into $p$. 
		Generically, for most points, the integral line passing through them will end at a minimum, forming a basin around this minimum. However, some of the integral line (starting from within a neighborhood of a maximum) will end at an index-($d-1$) saddle point, forming the separation between different basins around valleys. 
		These integral lines are exactly the union of stable manifolds of index-($d-1$) saddle points. They form a network of connections between mountain peaks (maxima) to saddles then to other mountain peaks, separating different valley basins around minima. See Figure \ref{fig:yusutmp} (b) for a 2D example. 

		We use the stable manifolds of the index-2 saddles of $f$ as the graph skeleton of the input density field $f: \mathbb{R}^3\to \mathbb{R}$. 
Intuitively, along a hidden neuron branch, the density values should be higher than points off the neuron branch. Using this terrain illustration, this means that intuitively the mountain ridges of this terrain (connecting mountain peaks to saddles then to neighboring mountain peaks) correspond to where hidden neuron branches lie, as off the mountain ridges, the function values will decrease (flow into different minima / valley basins). The 1-stable manifolds of the index-2 saddle points capture such mountain ridges.  

		\paragraph{Remark.} There is a dual concept of \emph{unstable manifold} $U(p) = \{p\}\cup \{x \in \R^d \mid ~ori(x) = p\}$ of a critical point $p$, consisting of all points along integral lines originated from $p$ (i.e, flowing away from $p$). In our case, for $d=3$, the stable manifold of a $p$ in $\rho$ is identical to the unstable manifold of $p$ in $-\rho$ (although the index of the critical point $p$ changes from $k$ to $3-k$, for $k = 0,1, 2, 3$).
		Computationally, as we will implement the above idea via discrete Morse theory in the discrete setting, it turns out that using the unstable manifolds of index-1 saddle points is much more efficient and simpler than using the stable manifolds for index-2 saddle points. Hence in our implementation, we will negate the density map and compute 1-unstable manifold for the function $-\rho: \myimage \to \mathbb{R}$ instead. Note that such 1-unstable manifolds connect minima to index-1 saddles to other minima and separate different mountain peaks for the map $-\rho$. 

\begin{figure}[htbp]
\begin{center}
\begin{tabular}{ccc}
	\includegraphics[height=4cm]{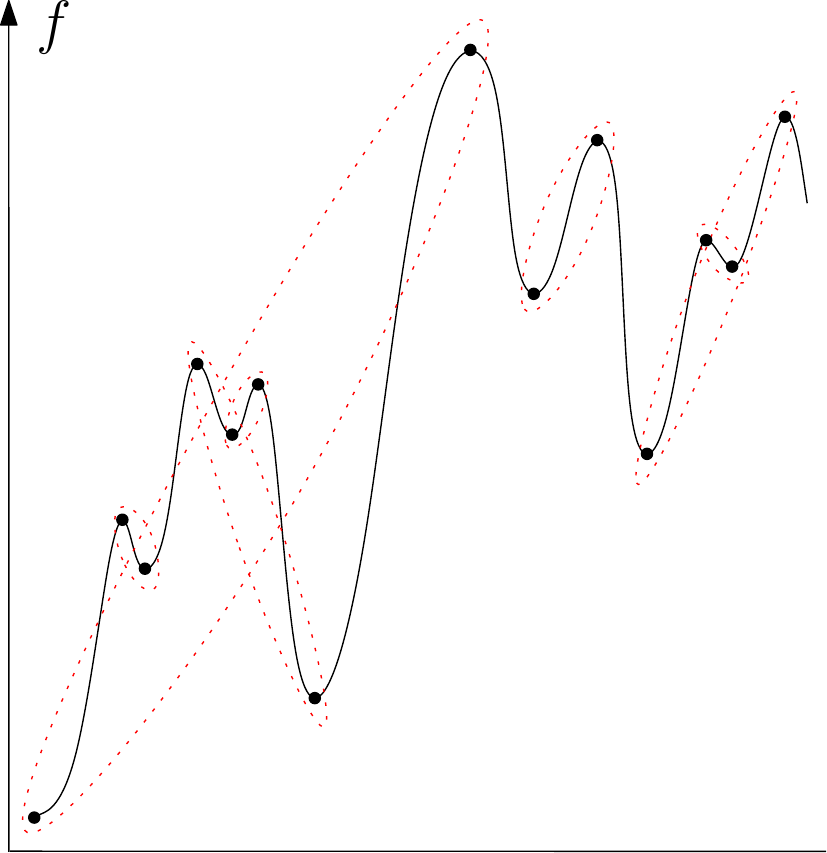} & &
	\includegraphics[height=4cm]{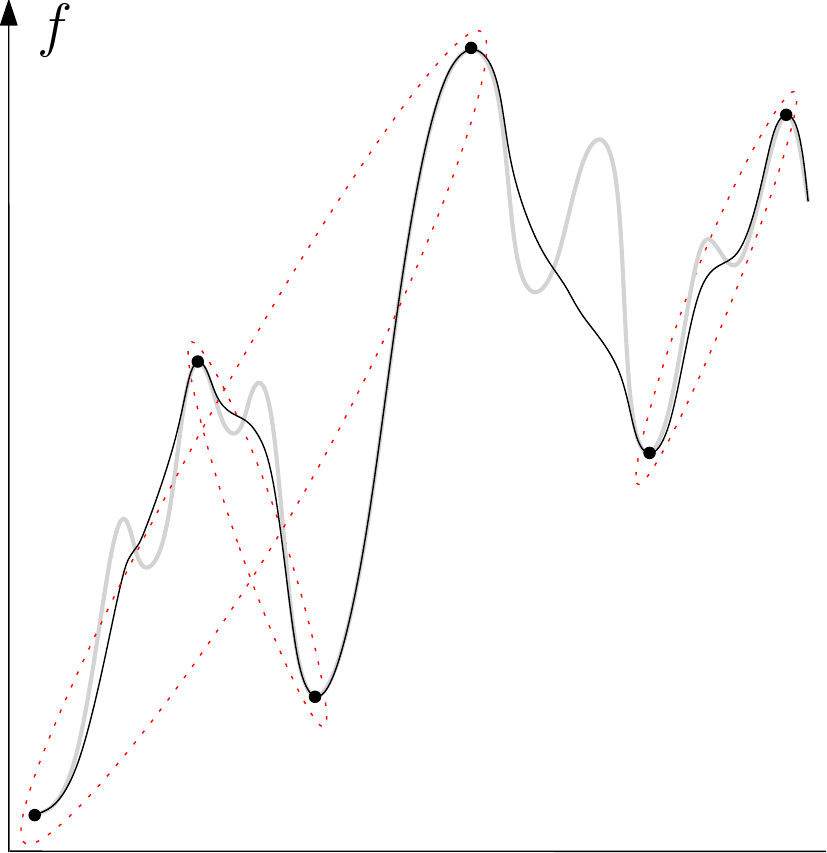} \\
(a) & & (b)
\end{tabular}
\end{center}
\caption{An example of the persistence pairing on a 1D function $f: \mathbb{R} \to \mathbb{R}$ as shown in (a): The persistence algorithm will pair up the critical points of function $f$ (as indicated by dotted ellipses), and the height difference between each pair is the persistence (importance) of the two critical points involved. 
In (b), we simplify those low-persistence critical points and only important peaks/valleys remain.
\label{fig:yusuNoise}}
\end{figure}

\paragraph{Noise removal.} Finally, note that the input images can be noisy, producing spurious critical points and thus spurious branches in the extracted graph skeleton (1-stable manifolds). Intuitively, we wish to identify such ``spurious'' critical points and ignore their corresponding 1-stable manifolds. 
To this end, we use the persistent homology, introduced in \cite{Edelsbrunner2002, zomorodian_2005}, to identify these critical points. In particular, using the persistent homology induced by the so-called lower-star filtration w.r.to the density map $\rho: \myimage \to \mathbb{R}$, one can obtain a \emph{persistence} value for each critical point\footnote{We note that persistent homology is one of the most important development in the field of topological data analysis in the past two decades, and has already been applied to a broad range of applications \cite{AEHW06, Carlsson:2004, chung, Edelsbrunner_persistenthomology,persistsurvey,Lamar, Liu2016ApplyingTP, platt} due to its power in feature characterization and quantification.}. 
This ``persistence'' indicates the importance of the critical point in a meaningful manner, which intuitively corresponds to the amount of perturbation in the input function $\rho$ one has to introduce in order to remove the (topological) feature introduced by this critical point. 
Critical points with small persistence are potentially caused by noise. Hence to remove noise, we will ``smooth'' these low-persistence critical points out, and consider only index-1 saddles whose persistence is larger than a given threshold $\tau \ge 0$, and output the union of 1-unstable manifolds of these saddles as the reconstructed graph skeleton $G \subset \myimage$. 
For simplicity of illustration, we show an example of persistence induced by a 1D function in Figure \ref{fig:yusuNoise}. 
In our pipeline, here we typically use a very low threshold to remove obvious noise without disconnecting the reconstructed skeleton. More simplification is performed later after we retrieve the tree summary from this graph skeleton in Step 3. 

\paragraph{Implementation in the discrete setting.} 
In practice, we are given a triangulation $\KK$ of the domain (3D cube) $\myimage \subset \mathbb{R}^3$, and the density function $\rho$ is given at the vertices of $\KK$. As mentioned earlier, our implementation only needs the so-called \emph{2-skeleton} of $\KK$, that is, the collection of vertices, edges and triangles. Our software \emph{DiMorSC($\KK, \rho, \tau$)} (DiMorSC stands for \emph{Discrete Morse on Simplicial Complex}) will take $\rho: \KK \to \mathbb{R}$ and a persistence threshold $\tau$ as input, and output the graph skeleton consisting of the 1-unstable manifolds for all index-1 saddles with persistence larger than $\tau$. 
Following \cite{Gyulassythesis,DRS15,2011MNRAS}, the computation of the 1-unstable manifold for this discrete setting is done through the use of discrete Morse theory, which is combinatorial in nature -- In particular, it only maintains the so-called \emph{discrete gradient vector field on $\KK$}, which consists of a collection of vertex-edge and edge-triangle pairs, and never approximates ``gradient vectors'' in a numerical manner. 

The high-level description of our discrete-Morse based skeleton extraction algorithm in Algorithm \ref{alg:dimorsc}. We note that this is similar to the \emph{persistence-guided discrete Morse based} graph reconstruction framework introduced in \cite{DWW17} which builds upon \cite{Gyulassythesis,DRS15,2011MNRAS} \footnote{The work of \cite{DWW17} builds upon, but simplifies both conceptually and implementation speaking, the work of \cite{Gyulassythesis,DRS15,2011MNRAS} for the specific case of extracting graph skeleton. However, those work are more general in the sense that they can compute higher dimensional (un)stable manifolds as well, while the framework formulated in \cite{DWW17} is specifically for 1-unstable manifolds and simpler.}. We thus will only provide a brief description below and for further details we refer the readers to \cite{DWW17} or to Chapter 5 and 6 of \cite{Suyithesis}. (We point out that our algorithm and its implementation in fact predates the work of \cite{DWW17}.)  
To our best knowledge, our software is currently the only publicly available code to extract the 1-unstable manifold from an arbitrary simplicial complex. 
In Section \ref{sec:stitching}, we will propose and incorporate a divide-and-conquer strategy into our software to handle large input images. 

\begin{algorithm}[h] 
	\caption{$G$ = DiMorSC($\KK$, $f$, $\tau$)}
		\label{alg:dimorsc}
		\begin{algorithmic}[1]
		\STATE Set $\hat{f} = -f$
		\STATE $\mathcal{P} =$ persistence pairing induced by lower-star filtration of $\KK$ w.r.t. $\hat{f}$
		\STATE $//$ ~$\mathcal{P}$ consists a set of vertex-edge and edge-triangle pairs
		\STATE Initialize the discrete gradient vector field $\disV$ on $\KK$

		\FOR{each vertex-edge pair $\langle v, e\rangle \in \mathcal{R}$ s.t. $per(v,e) \le \tau$} 
			\STATE {Cancel (simplify) this pair $\langle v, e\rangle$ and update the discrete gradient vector field $\disV$}
		\ENDFOR

		\STATE $G = \emptyset$
		\FOR{each critical edge $e$ with $per(e) > \tau$} 
			\STATE $G = G \cup \{$ 1-unstable manifold($e$) in $\disV \}$
		\ENDFOR
		\RETURN $G$
		\end{algorithmic}
	\end{algorithm}

In Algorithm \ref{alg:dimorsc}, line 2 computes the persistence pairing to assign importance (persistence) of critical simplices (analogous to critical points in the smooth setting). The computation is done using the library PHAT \cite{BAUER2017}, which provides state-of-the-art performance in computing persistence homology. 
Lines 3--7 simplify the discrete gradient vector fields by canceling (thus destroying) critical simplices with low persistence. This is analogous to the ``smooth-out'' of low-persistence critical points in the continuous setting as illustrated in Figure \ref{fig:yusuNoise}. Finally, lines 8--11 collect the 1-unstable manifolds for high-persistence critical edges (corresponding to index-1 saddle points) as the graph skeleton of density field $\rho$. The implementation details of Lines 3 -- 11
can be found in the PhD dissertation of one of the co-authors, Suyi Wang \cite{Suyithesis}.

	\subsection{Step 3: Summary tree extraction and simplification}
	\label{subsec:treesimp}
		The output of the above Morse-based skeletonization step is a geometric graph $G$ and we will convert the graph into a (simplified) tree $T$ as the summarization of the input 3D image $\myimage$. We provide a further simplification strategy to control the level details of the final summary tree keeping in mind the application to skeletonizing tracer injections, so that skeletons that capture more or less detail can be produced. 

\begin{figure}
\centering
\includegraphics[width=\textwidth]{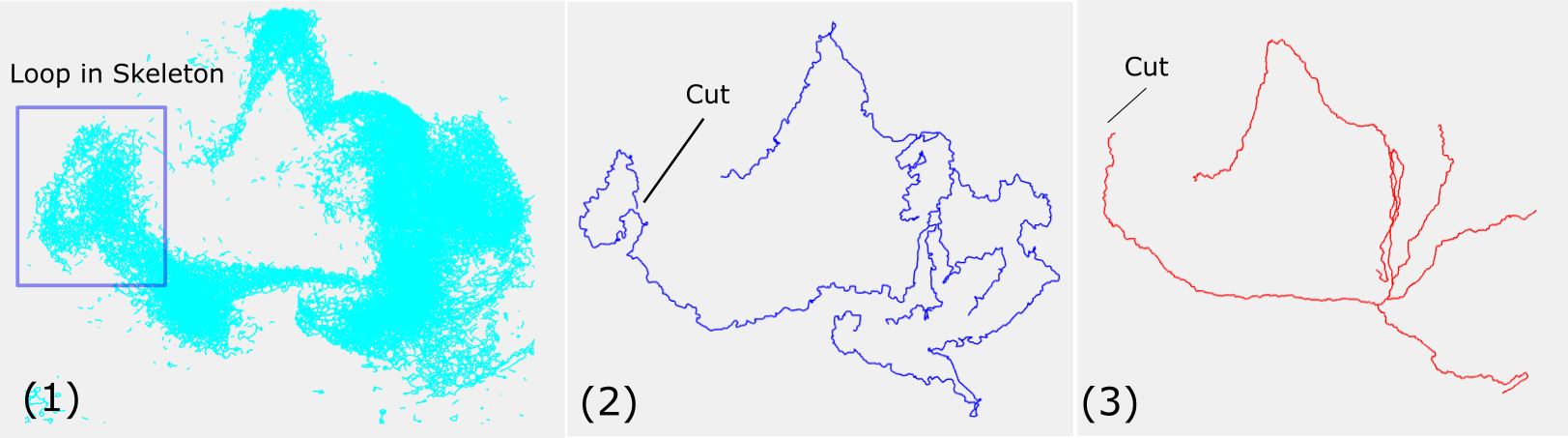}
\caption{A comparison between the output tree extracted (and simplified for more clear view) via the minimum spanning tree approach, shown in (2), and via the shortest path tree approach, shown in (3); generated from the same input as shown (1). The two approaches resolve loop differently: a loop is always cut at a location locally furtherest away from the root for the shortest path tree, while there is less control of how the loop is cut for the minimum spanning tree.
\label{fig:mst}}
\end{figure}

\begin{wrapfigure}{r}{0.4\textwidth} 
\includegraphics[width=0.3\textwidth]{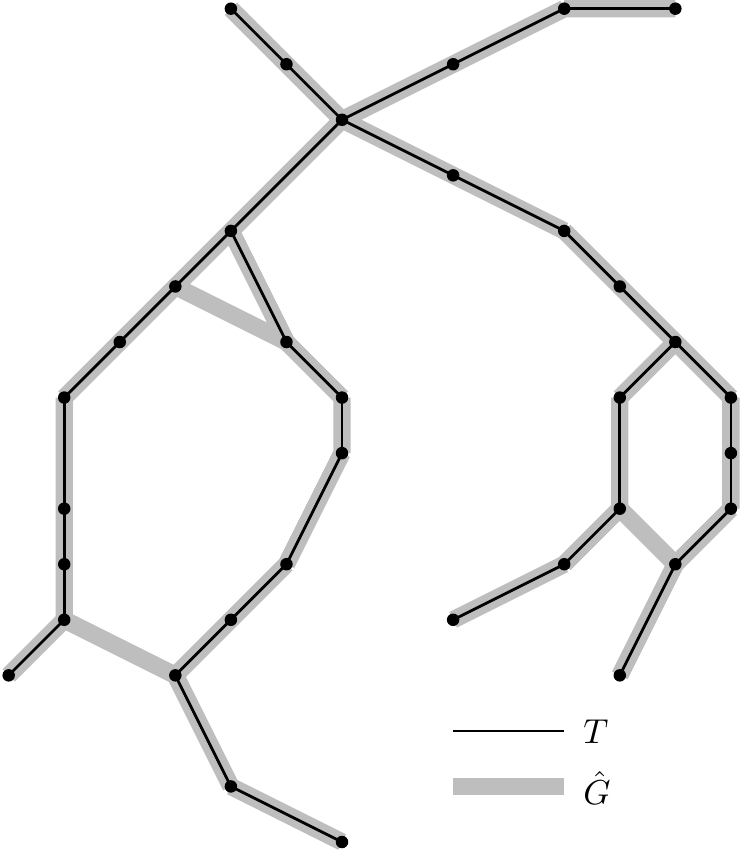}
\caption{
\label{fig:To_tree}}
\end{wrapfigure}

		\paragraph{Extraction of a rooted tree.}
		Our input is the graph skeleton $G$ extracted in {\sf Step 2}. Note that each arc $e\in E$ in graph $G = (V, E)$ is realized by a polygonal path, consisting of edges from the input triangulation $\KK$. 
We now augment $G$ to $\Ghat=(\Vhat, \Ehat)$ so that its node set $\Vhat$ includes both those from $V$ and all vertices of the edges from each arc in $E$. See Figure \ref{fig:To_tree} for an illustration where black dots are vertices in $\Vhat$. 
		We now extract a rooted tree $T$ from $\Ghat$, which is in fact a spanning tree of $\Ghat$ (that is, $T$ contains all vertices in $\Vhat$, and edges of $T$ are from $\Ehat$). Note that here we assume that $\Ghat$ is connected -- if not, we will extract a tree summary for each connected component of $\Ghat$. 

		We also assume that we are given the choice of the tree root $\myroot$ -- in our experiments, this is typically set to be the injection site, which is easy to infer due since the neuronal somata are concentrated at the injection site. In general, the choice of the root $\myroot$ can also be taken as the soma of a neuron in the single neuron reconstruction. 
		We then take the shortest path tree w.r.to source $\myroot$ as the initial tree summary $T$, where the length of a path is measured by the number of edges in it. (That is, we assume that all edges in graph $\Ghat$ has weight 1.) 
		It is possible to weight the edge by a quantity proportional to the inverse of the density $\rho$ along this edge. Our software provides this option. However, our current choice appears to work well in current experiments. 

		Note that it may also be possible to use the maximum spanning tree of $\Ghat$ (where the weight of each edge $(u,v)\in \Ehat$ equals $(\rho(u) + \rho(v))/2$). This strategy also makes sense as it encourages to include high density edges in the spanning tree. 
Indeed, we tested both methods in our experiments. The two strategies often generate similar results, although we observe that shortest path tree strategy tends to produce more natural trees while maximum-spanning-tree strategy sometimes introduces breaks in the middle of a long branch. Specifically, when a branch contains relatively weak signal (noise) in the middle resulting in its end point being mis-connected to other branches and thus forming a loop, making the maximum spanning tree more likely to cut the branch in the middle. In contrast, the shortest path tree (starting from a reliable choice of root) would cut the branch in the far end, which is more consistent to neuron morphology. An example is shown in Figure \ref{fig:mst}. In cases when there is no obvious choice of tree root, we have utilized the maximum spanning tree strategy. 
		
		\paragraph{Tree simplification.}
		The initial tree $T $ constructed above may contain an excess of detail. We further develop the following simplification strategy to allow the users to control the desired level of detail in the tree summary. 

		Specifically, given the tree $T=(\Vhat, E_T)$ rooted at $\myroot$, we first assign a function $\geo: \Vhat \to \mathbb{R}$ for all nodes in $\Vhat$ by, for any vertex $v \in \Vhat$, 
$$\geo(v) = \sum_{e\in \pi(\myroot, v)} {length(e) * weight(e)}, $$
where $\pi(\myroot, v)$ is the unique tree path from the root $\myroot$ to node $v$. 
The length of an edge $e$, $length(e)$, is simply the Euclidean distance between the two endpoints of $e$.
As for the weight of $e$, i.e, $weight(e)$, we provide two choices: (i) a uniform weight where $weight(e) = 1$ for all edges $e\in E_T$; and (ii) an intensity-based weight $weight(e) = (\rho(v_1) + \rho(v_2)) / 2$, where $v_1, v_2$ are adjacent vertices of edge $e$.	
		Note that the function $\geo$ is monotonically increasing along any root-to-leaf path in the tree $T$, and $\geo(\myroot) = 0$. 

		Let a \emph{leaf node} refer to any degree-1 node that is not the root, and a \emph{junction node} denote any node with degree larger than 2. 
		We now describe a natural \emph{branch decompositon} procedure that partitions the tree $T$ into a set of branches, and also assigns a measure of importance to each branch. During the simplification process, we simply remove those branches with small ``importance''. Interestingly, this decomposition as well as the importance assigned to each branch are exactly the information encoded in the persistent homology induced by the so-called \emph{super-level set filtration} w.r.to the function $\geo$ (viewed as a piecewise-linear function on $T$). This relation holds as the function $\geo$ is monotone along any root-to-leaf path in $T$. Hence our simplification procedure essentially removes those branches (topological features) less important under persistent homology w.r.t. $\geo$. While we point out this connection to persistent homology here, in what follows, we will only describe the branch decomposition / persistence-assignment procedure. 

\begin{wrapfigure}{r}{0.4\textwidth} 
\centering \includegraphics[width=0.3\textwidth]{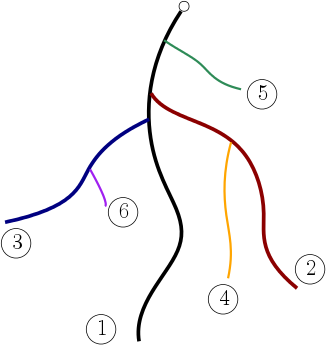}
\caption{Branch decomposition of a rooted tree, where for simplicity assume $g(v)$ simply equals to distance (along the tree) to the root (white dot). The color paths are the decomposed branches, with indices indicate their persistence order (so number 1 indicates the branch with largest persistence). 
\label{fig:treedecomposition}}
\end{wrapfigure}
		Let $L = \{\ell_1, \ell_2, \ldots, \ell_t\} \subset \Vhat$ denote the leaf set. 
		We will partition $T$ into a set of $t = |L|$ branches $\Pi =\{\pi_1, \ldots, \pi_t\}$, each each branch $\pi_i$ is a path from the leaf node $\ell_i$ to a junction node or to the root. The union of paths in $\Pi$ equals $T$, while all branches are disjoint other than potentially at their two end points. 
		
        Specifically, at the beginning, we have a single tree $T$ with root $\myroot$. Let $\ell_1$ be the leaf node with \emph{largest} $\geo$ function value, and let $\pi_1$ be the unique path from $\myroot = root(T)$ to $\ell_1$. 
		To continue, remove path $\pi_1$ from $T$, which will decompose $T$ into a set of disjoint trees $T_1, \ldots, T_k$, the root of each of them will necessarily be a junction node in $T$. If $k = 0$, then this process terminates. 
		Otherwise, we repeat the same procedure for each tree $T_i$, $i\in [1, k]$, recursively; and let $\Pi$ denote the collection of all the branches obtained along the way. See Figure \ref{fig:treedecomposition} for an example. 
        
		After we compute the branch decomposition $\Pi$, we assign the persistence (importance) of each branch $\pi_i$ as follows: Let $\ell_i$ and $s_i$ be the two endpoints of $\pi_i$ where $\ell_i$ is a leaf node and $s_i$ is either a junction node or the root $\myroot$ of $T$. We set $per(\pi_i) = per(\ell_i) = per(s_i) = \geo(\ell_i) - \geo(s_i)$. 
		As said earlier, it turns out that the set of pairings $\{(\ell_i, s_i) \mid i\in [1, t] \}$ is exactly the set of persistence pairing produced by the persistent homology induced by the super-level set filtration of $\geo$, and the persistence of the corresponding branch equals to the persistence of the pair $(\ell_i, s_i)$. 
		
\begin{figure}
	\centering
    \includegraphics[width=0.9\textwidth]{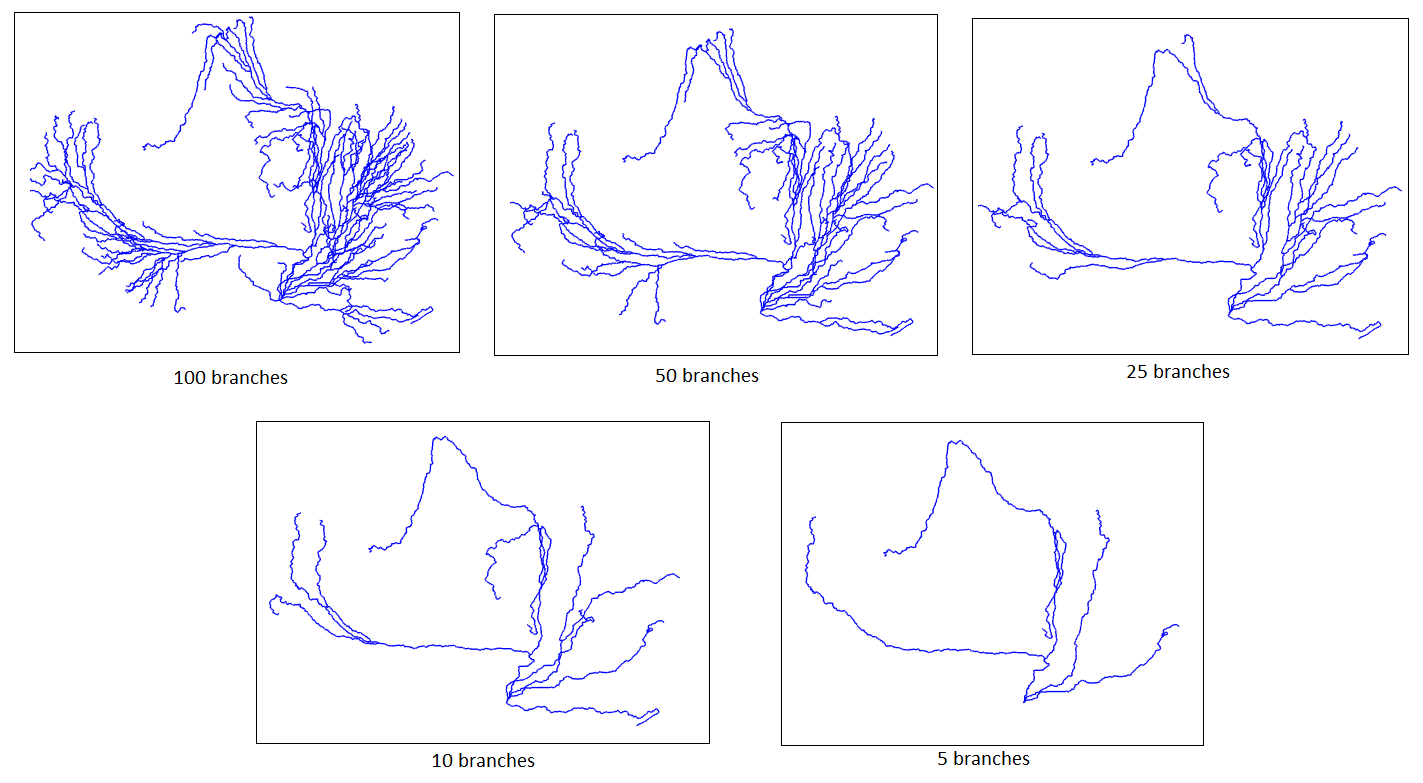}
	\caption{Simplification with branches of 100, 50, 25, 10 and 5, respectively.}
	\label{fig:treesimp}
	\end{figure}

		Given a threshold $\tau$, to remove unnecessary details in $T$, we simply output the subtree $T_\tau$ formed by the union of branches from $\Pi$ with persistence larger than $\tau$. It is easy to show that the union of these branches is necessarily connected (i.e, a single tree) if the input tree $T$ is connected. See Figure \ref{fig:treesimp} for examples. 

\section{Divide-and-conquer strategy for handling large images}
\label{sec:stitching}
        \begin{figure}
        \begin{center}
        \begin{tabular}{ccc}
        \includegraphics[width=0.45\textwidth]{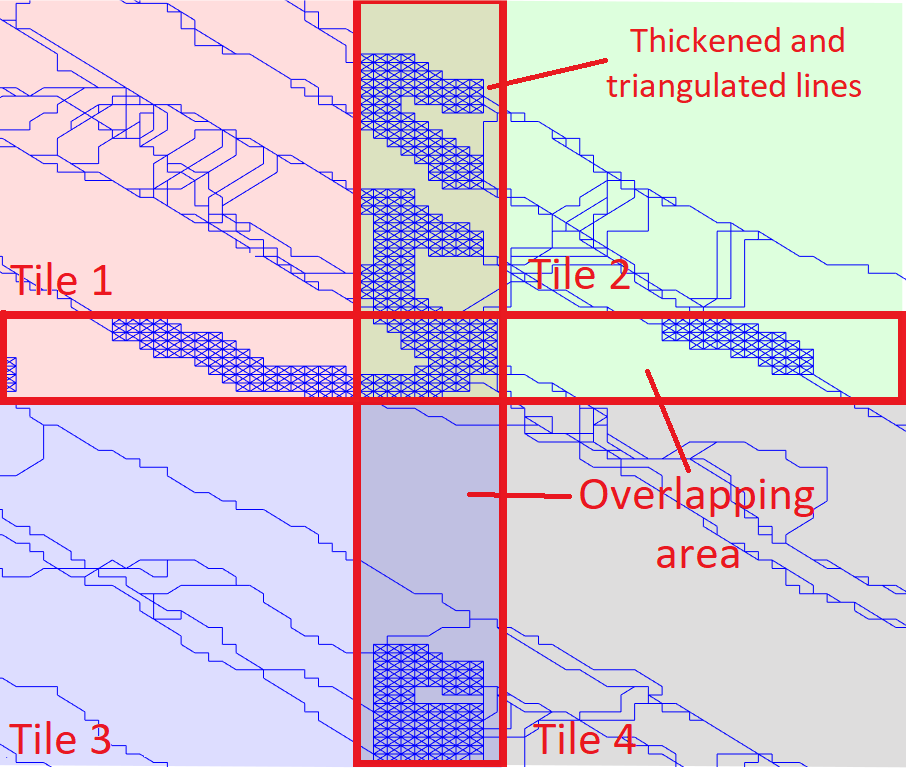} & \hspace*{0.1in} &
        \includegraphics[width=0.45\textwidth]{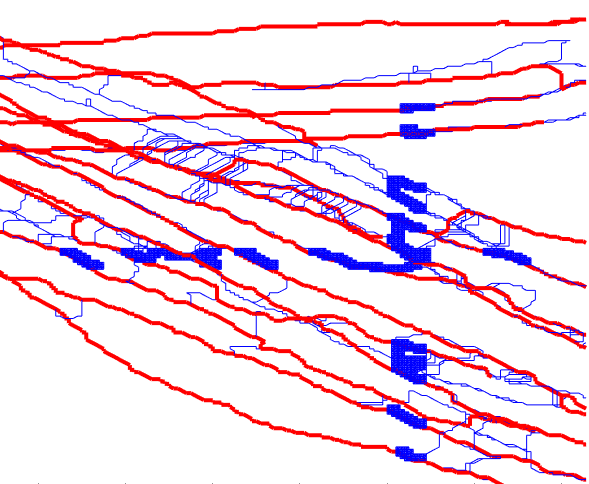}\\
        (a) & & (b)
        \end{tabular}
        \end{center}
        \caption{(a) The original domain is divided into four tiles (different colors). The blue graphs are the reconstruction within each tile, and the blue cells are those in $\KK_{merge}$. 
        (b) Blue curves/cells are $\KK_{merge}$ and the red graph is the merged graph and then simplified.
        }
        \label{fig:merge}
        \end{figure}
The tracer-injected data sets have around 300 images of brain sections in each brain volume. Each of these section images have $18000*24000$ pixels creating a rather large dataset.
		This type of data is too large to fit into memory and to be processed all together at the same time.  Hence we have developed the following divide-and-conquer approach to handle such large images.

		In the high level, our approach (i) partitions the input image stack $\myimage$ into 
		$k$ small tiles (cubes) $\tile_i, i \leq k$, 
		(ii) extracts graph skeleton $\stable_i$ in tile $\tile_i$ and 
		(iii) merges all the results $\stable_i$ into a single graph skeleton $\stable$.
		
		In our experiments, the size of each tile is chosen to be $512 * 512$ in xy-direction
		and we have not partitioned in the z-direction as the number of slices is typically only a few hundreds. 
		(However, one can easily perform partitioning in z-direction as well if the needs arise.)  
		This tile size represents a good trade off between information encoded in each tile and
		the processing time.  we have also set a small overlapping area (5 pixels) between adjacent tiles.
		
See Figure \ref{fig:merge} for an example, where different colors indicate different tiles and they share an overlapping area.
		Then the 1-stable manifold is extracted for each tile using Algorithm \ref{alg:dimorsc} presented in the previous section.
		The key step is (iii), the merging of the graph skeletons $\stable_i$s into a single graph $\stable$ representing the skeleton of the merged tiles. 
		We describe how to perform the merging step now. 

		Let $\KK_i \subseteq \KK$ denote the subtriangulation of tile $\tile_i$, $i\in [1, k]$. 
		Let ${V_i}$ be the set of internal vertices for $\tile_i$ and
		let ${V^b_i}$ be the set of vertices in 
		all overlapping area, $\bigcup_{j} \tile_i \cap \tile_j$, between $\tile_i$ and any neighborhing tile $\tile_j$. 
		Recall that the graph skeleton $\stable_i$ is extracted from the density function $\rho_i : \KK_i \to \R$, which is the restriction of $\rho$ to $\KK_i$.  
		Let ${E_i}$ be the set of edges in ${\stable_i}$.
		
		To merge $\stable_i$s in a natural manner, we will leverage the 1-unstable manifold based framework again: 
		First, we will build a new simplicial complex $\KK_{merge} \subseteq \KK$ as follows:
		For each $i\in [0, k]$, we take a small neighborhood (a small triangulated cubic region) around every vertex $v \in V^b_i$; let $C_i$ denote the union of such triangulated neighborhoods for all vertices in $V^b_i$. 
		We set $\KK_{merge} = \bigcup_i \left(\stable_i \cup {C_i}\right)$; see Figure \ref{fig:merge} (a). 

		Next, we diffuse the function values of $\rho(v)$, for all $v\in \bigcup_i V^b_i$, to vertices in $\KK_{merge}$ using a Gaussian kernel, and establish a function $\rho': \KK_{merge} \to \mathbb{R}$ on $\KK_{merge}$. 
		In other words, for any $u \in \KK_{merge}$, $\rho'(u) = \sum_{v\in \cup_i V^b_i} e^{-\frac{\|v - u\|^2}{2\sigma^2}} \rho(v)$, where 	$\sigma^2$ is the variance of the Gaussian kernel.

		Finally, we perform Algorithm \ref{alg:dimorsc} on $\rho': \KK_{merge}\to \mathbb{R}$, and extract the final merged graph $\stable$.

		Note that compared to the size of each tile $\tile_i$, the extracted 1-skeletons $\stable_i$ from it
		is much smaller (as it intuitively only represents potential signals in input image). Hence the size of $\KK_{merge}$ is potentially far smaller than the size of $\KK$, and 			the memory cost for processing an individual tile as well as the merging process is controlled at a
		reasonable amount, allowing our algorithm to handle large data not fit in memory.
		This approach is only a small modification of the original pipeline, but can greatly increase the size of data it can handle.


\begin{figure}
    \centering
    \includegraphics[width=\textwidth]{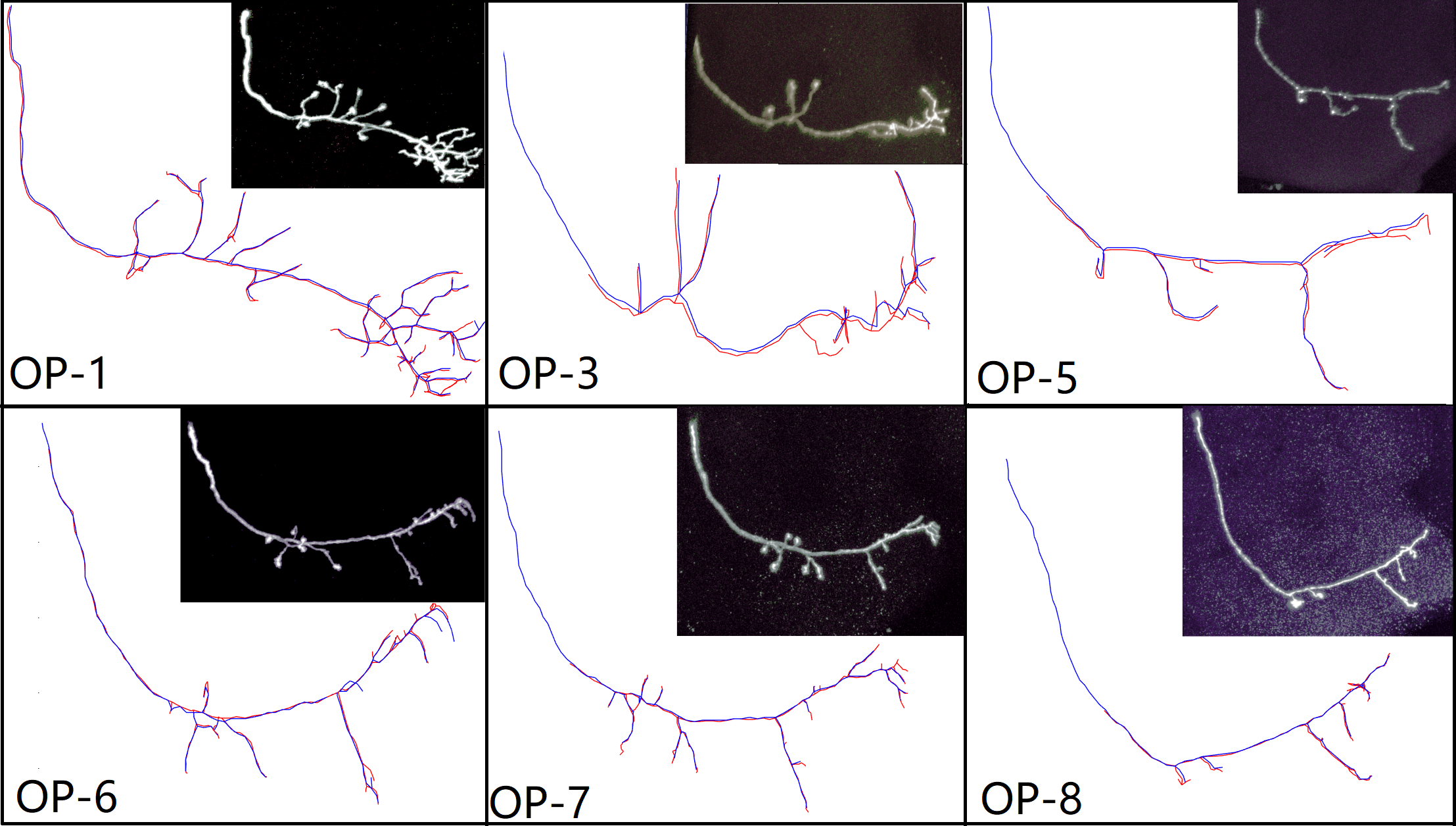}
    \caption{The result (blue) superpose on ground truth (red) of Olfactory Projection dataset in DIADEM challenge \cite{diademchallenge}. }
    \label{fig:OP_result}
\end{figure}

\section{Results}
	\label{sec:result}

\begin{figure}[htbp]
\begin{center}
\begin{tabular}{ccc}
\includegraphics[height=4cm]{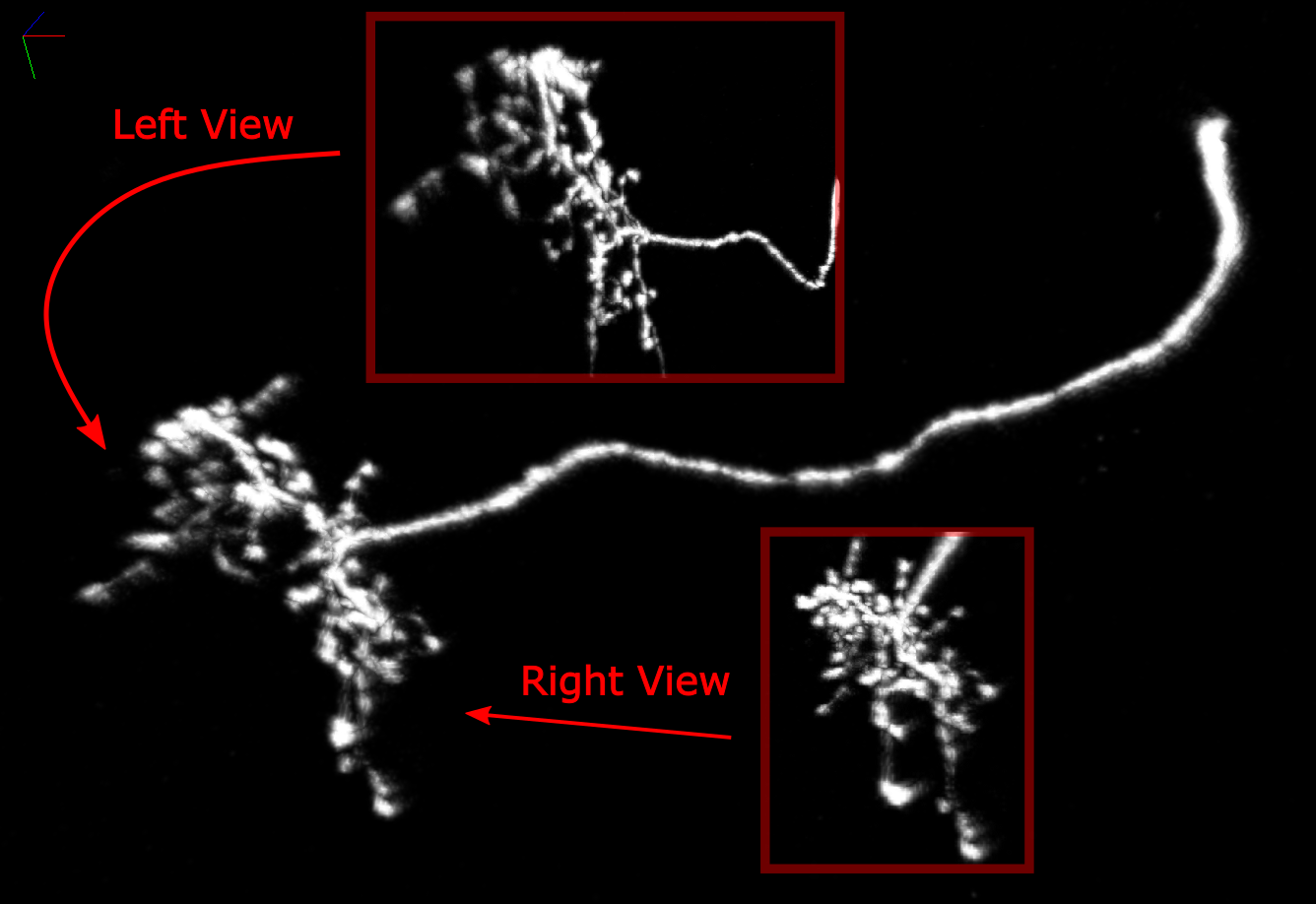} & \hspace*{0.1in} &
\includegraphics[height=4cm]{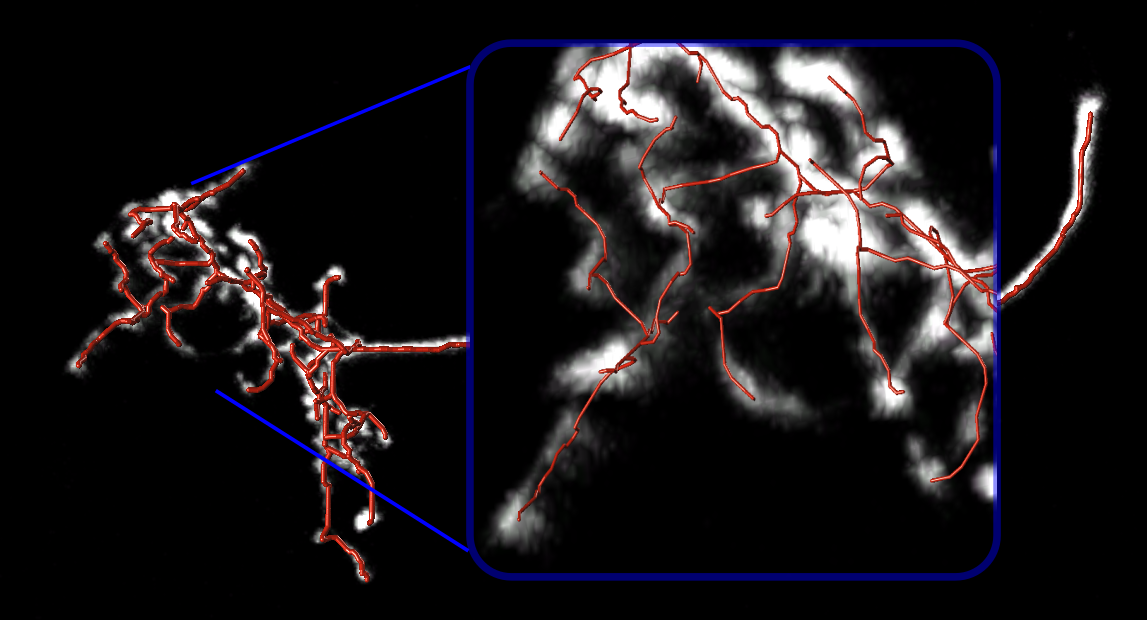}\\
(a) & & (b)
\end{tabular}
\end{center}
\caption{(a) The input signal has non-uniform strength with gaps (the dataset is OP-9 from DIADEM dataset). However, the reconstruction (as shown in (b)) based on global topological structure of the signal density field can still connect through them. 
\label{fig:yusuOP9}}
\end{figure}

	\subsection{Single neuron tracing}
    \label{subsec:exp:single}
    
    As proof of principle demonstration of our proposed pipeline, we first show the performance of our discrete Morse based framework in reconstructing neurons from the 
    Olfactory Projection Fibers dataset (OP dataset) provided as part of the DIADEM challenge \cite{diademchallenge}. 
    The dataset contains nine stacks of drosophila olfactory axonal fibers in Olfactory Bulb and the roughly 512*512*60 resolution images are acquired by 2-channel confocal microscopy method.
    This dataset is accompanied by manual tracing results which can serve as ground truth, allowing for qualitatively comparison of the output of reconstruction with the ground truth, called DIADEM score \cite{diadem_metric}. Below, we show our reconstruction visually, and compare them qualitatively with other popular methods using the DIADEM score.

\begin{figure}[htbp]
\begin{center}
\begin{tabular}{ccc}
\includegraphics[height=4cm]{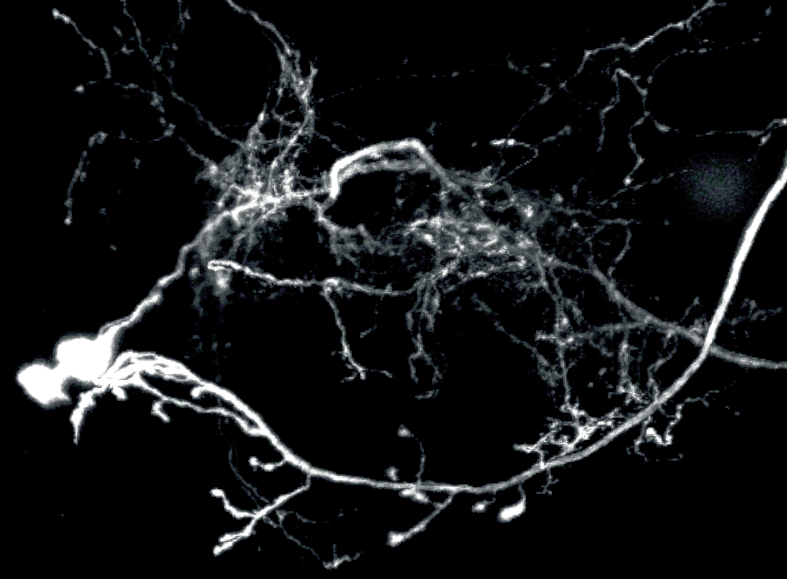} & \hspace*{0.1in} &
\includegraphics[height=4cm]{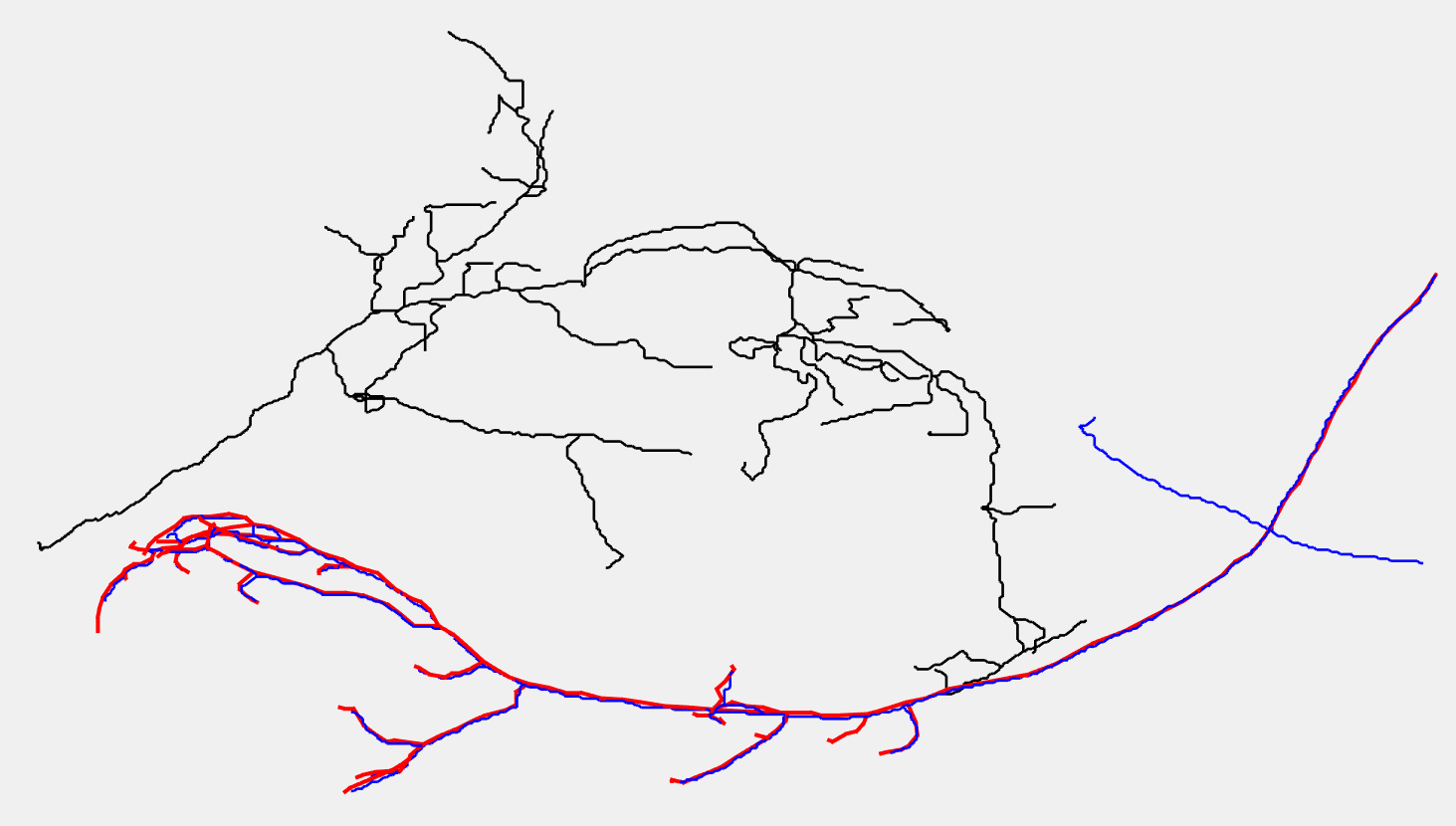}\\
(a) & & (b)
\end{tabular}
\end{center}
\caption{In the data set OP-2 from DIADEM, there are actually two neurons shown in the image, with the ground truth given for the front one. Our method reconstructs both neurons. 
\label{fig:yusuOP2}}
\end{figure}

  Figure \ref{fig:OP_result} shows the tracing results of DiMorSC for OP dataset
    No. 1, 3, 5, 6, 7, and 8. Datasets OP-2 and OP-9 will be reported later in Figure \ref{fig:yusuOP9} and \ref{fig:yusuOP2}. 
    As we can see that even though the input signal may not always uniform (see Figure \ref{fig:yusuOP9} for a detailed example on dataset No. 9 referred to as OP-9), as our method relies on the global structure of the signal density field, it still connect through these gaps. 
    We also show the input and reconstruction of OP-2 data set in Figure \ref{fig:yusuOP2}, where as we can see that interestingly, there are actually 
   two neurons in this image. The ground truth is given only for the front one. Our method can reconstruct both neurons as shown in blue and black colors. 

 \begin{table}[htbp]
    \centering
    \begin{tabular}{ c | c | c | c | c}
    \hline			
    Set & DiMorSC & APP2 & Smart tracing & SNAKE \\
    \hline
    1 & 	0.914 & 	0.796 &	0.853	& 	0.827\\
    3 & 	0.765 & 	0 	&	0.605	&	0.766\\
    4 & 	0.804 &	0.727 &	0.79	&	0.704\\
    5 & 	0.755 & 	0.389 &	0	&	0\\
    6 & 	0.858 & 	0.857 &	0.779	&	0.667\\
    7 &	0.923 & 	0.773 &	0.906	&	0.788\\
    8 & 	0.849 &	0.373 &	0.696	&	0.725\\
    9 & 	0.833 & 	0.786 &	0.739	&	0.657\\
    \hline 
    \end{tabular}
    \caption{DIADEM scores for various OP datasets. 
    \label{fig:OP_score}}
    \end{table}
    
    \begin{table}[htbp]
    \centering
    \begin{tabular}{ c | c | c | c | c | c | c | c}
    \hline			
    Set &  Data  	& DM		& DM 	& DM 	& APP2	& Smart tracing & SNAKE \\
    & Loading & Step1 		& Step2 	& Step3  & 		& 				& 		\\
    \hline
    1 & 	0.67 & 	8.6 		& 19.07	&	0.37& $<$1s & 	6min		&		24\\
    3 & 	0.70 & 	5.51 		& 12.7	&	0.26& $<$1s	& 		8min	&	-	\\
    4 & 	0.73 &	6.76 		& 16.51	&	0.35& $<$1s	& 		8min13s	&	-\\
    5 & 	0.83 & 	4.85 		& 6.99	&	0.24& $<$1s	& 		6m15s	&	-\\
    6 & 	1.02 & 	5.60 		& 6.49	&	0.29& $<$1s	& 		4m07s	&	35\\
    7 &	0.71 & 	5.08 		& 7.94	&	0.28& $<$1s	& 		6m42s	&	20\\
    8 & 	0.95 &	7.56 		& 13.28	&	0.30& $<$1s	& 		13min13s&	28\\
    9 & 	0.96 & 	6.74 		& 10.89	&	0.33& $<$1s	& 		2m55s	&	31\\
    \hline
    \end{tabular}
    \caption{Running time in seconds for various OP datasets. ``DM'' stands for our DiMorSC algorithm, and columns 3--5 show the running time for step 1, 2, and 3 respectively.}
    \label{tab:OP_time}
    \end{table}
    For quantitative evaluation, in Table \ref{fig:OP_score} we report the \emph{DIADEM score} of our algorithm and that of the output by APP2 \cite{app2}, SmartTracking \cite{smarttracing} and SNAKE \cite{Wang2011} algorithms (which are three state-of-the-art algorithms in single neuron tracing). The DIADEM score \cite{diadem_metric} is a metric in range (0,1) 
    that measures the similarities between reconstruction and ground truth, where the higher the value is the more similar they are. This score has been used to evaluate the algorithms in the DIADEM challenge. 
    As we can see, we obtained similar or better DIADEM score in all datasets.
    We note that the DIADEM score could be sensitive to 
    both the root location and the geometric information of branches.
    For example, simply smoothing our result sometimes increases the DIADEM score,
    however, the smoothed branches are visually less aligned with the ground truth.
    The root location provided in the data sometimes lies obviously 
    in the middle of a long branch. 
    
    we also report the running time of our algorithm, as well as for the three algorithms mentioned above in Table \ref{tab:OP_time}. 
    In general, our algorithm finishes in 15s for the OP dataset, 
    which is slower than that of APP2 ($<1s$), comparable to SNAKE (average 25s) and
    faster than Smart tracing (average 6 min).
    However, we note that this is a simple implementation of our algorithm and we believe that it can be further improved. Indeed, we note that recent observations in \cite{DWW18} can significantly simplify our algorithm and improve its time efficiency.

    \begin{figure}
	\centering
	\begin{subfigure}{0.45 \textwidth}
	  \includegraphics[width=0.9\textwidth, height=5cm]{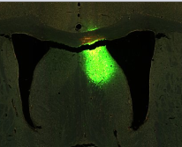}
	\quad
	\end{subfigure}
	\begin{subfigure}{0.45 \textwidth}
	  \includegraphics[width=0.9\textwidth, height=5cm]{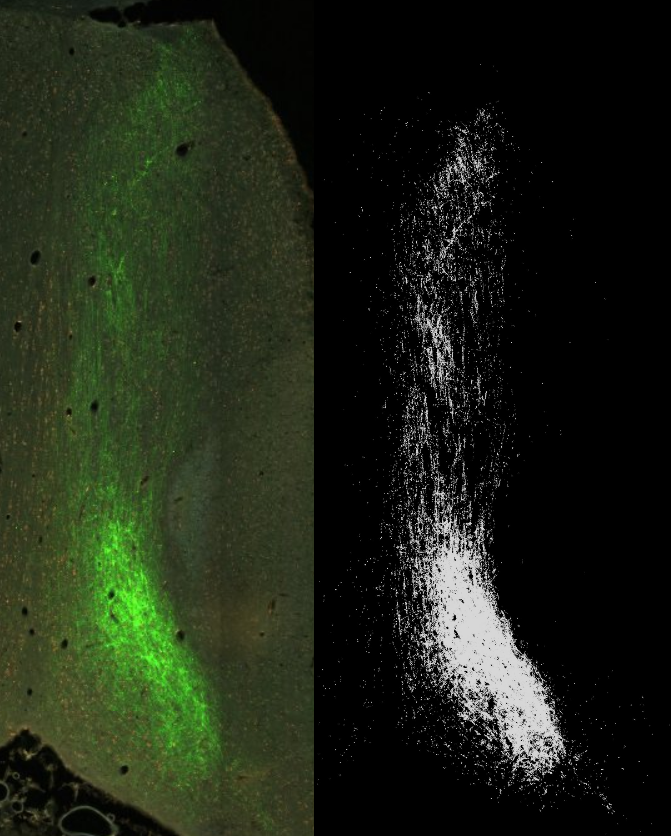}
	\quad
	\end{subfigure}
	\caption{LSc injection (in the left image) and signal detection (right images)}
	\label{fig:injection}
	\end{figure}

     \begin{figure}
	 \centering
	 \includegraphics[width=0.9\textwidth, height=9cm]{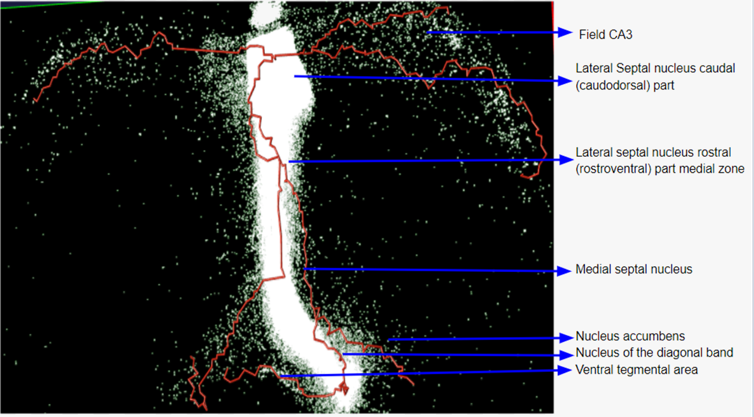}
	
	\caption{Skeletonization result on the preprocessing data}
	\label{fig:LScanno}
	\end{figure}
    
    \begin{figure}
		\centering
		\begin{subfigure}{0.32 \textwidth}
			\includegraphics[width=0.9\textwidth, height=4cm]{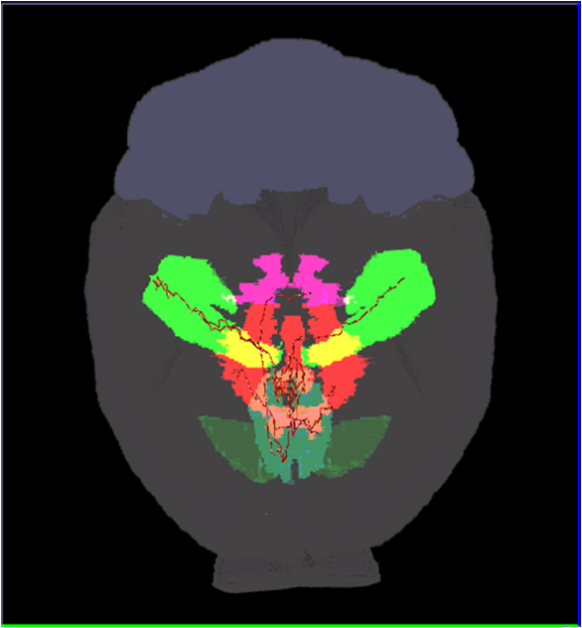}
			\quad
		\end{subfigure}
			\begin{subfigure}{0.32 \textwidth}
			\includegraphics[width=0.9\textwidth, height=4cm]{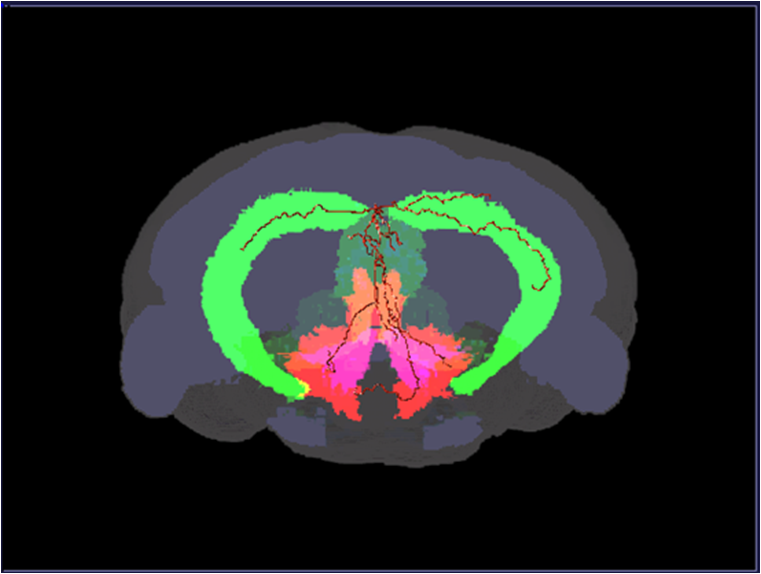}
			\quad
		\end{subfigure}
		\begin{subfigure}{0.32 \textwidth}
			\includegraphics[width=0.9\textwidth, height=4cm]{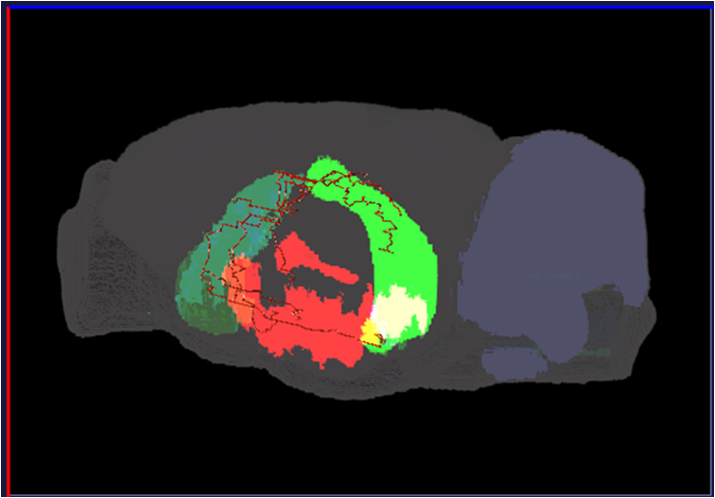}
		\end{subfigure}
		\caption{Neuron connectivity atlas}
		\label{fig:PMD_atlas}
	\end{figure}
   
\subsection{Mesoscopic summarization}

We also report our results anterograde tracer injected data 
    from the Mouse Brain Architecture Project\cite{PMDmouse}, which will be referred to as an 'MBAP' brain.
    The MBAP brain dataset is obtained from whole mouse brains where specific brain areas have been injected with an AAV florescent tracer. The example brain dataset contains a stack of about 270 images of brain sections with 18000*24000 pixels per section. 
    In every image slice (2D image), each pixel is 0.46 microns in both dimensions and 
    vertically, the distance between two consecutive image slices is 20 microns. In the example below, a brain with a single injection (green fluoroscence) in the brain area, the Lateral septal nucleus-caudal part(LSc) is shown.
    In addition to the large size of the data and the anistropic sampling, 
    there is background florescence in the image. 
    Although it may be possible to identify the neuron signal visually in the images by careful inspection, 
    automatic tracing of the trajectories of bundles of axons presents a significant challenge.

	The specific pre-processing procedure described in section 2.2 is applied to remove the background. 
    See Fig \ref{fig:injection} for fluorescence injection region and preprocessing result. 
 

	\begin{figure}
	\centering
	\begin{subfigure}{0.45 \textwidth}
	  \includegraphics[width=0.9\textwidth, height=5cm]{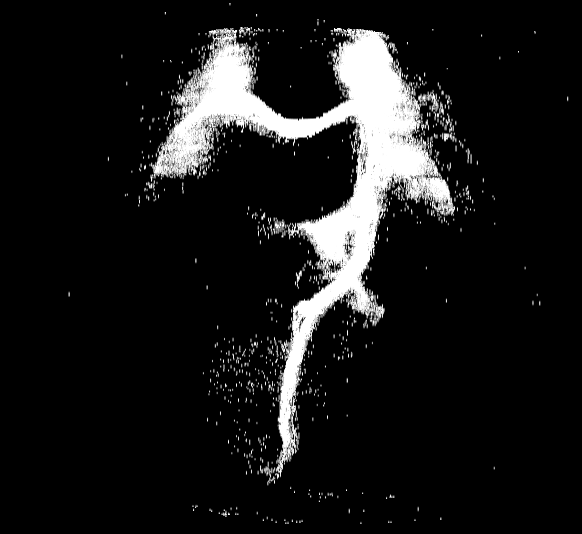}
	\quad
	\end{subfigure}
	\begin{subfigure}{0.45 \textwidth}
	  \includegraphics[width=0.9\textwidth, height=5cm]{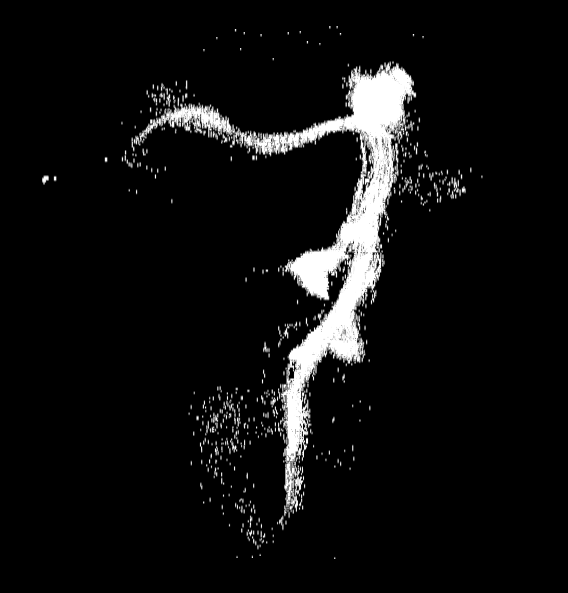}
	\end{subfigure}
    \begin{subfigure}{0.45 \textwidth}
	  \includegraphics[width=0.9\textwidth, height=5cm]{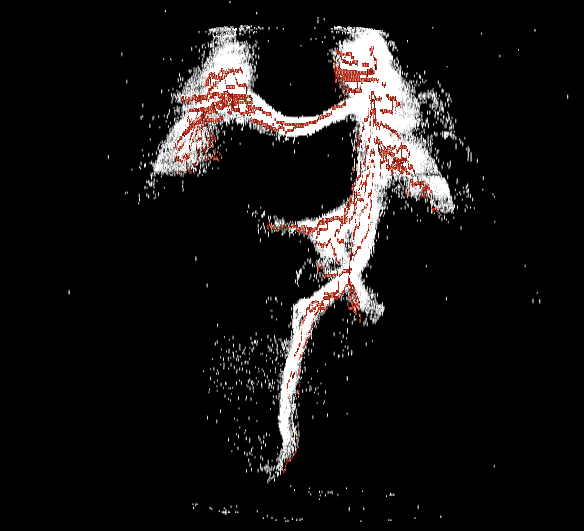}
	\quad
	\end{subfigure}
	\begin{subfigure}{0.45 \textwidth}
	  \includegraphics[width=0.9\textwidth, height=5cm]{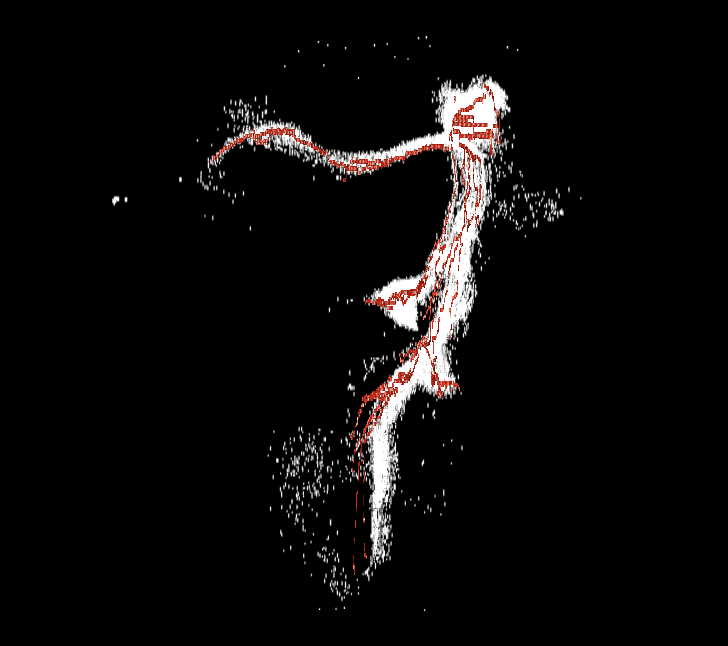}
	\end{subfigure}
	\caption{Motor Cortex injection and summarization}
	\label{fig:motorcortex}
	\end{figure}
    
    \begin{figure}
	\centering
    \begin{tabular}{ccc}
     \includegraphics[width=0.46\textwidth, height=6cm]{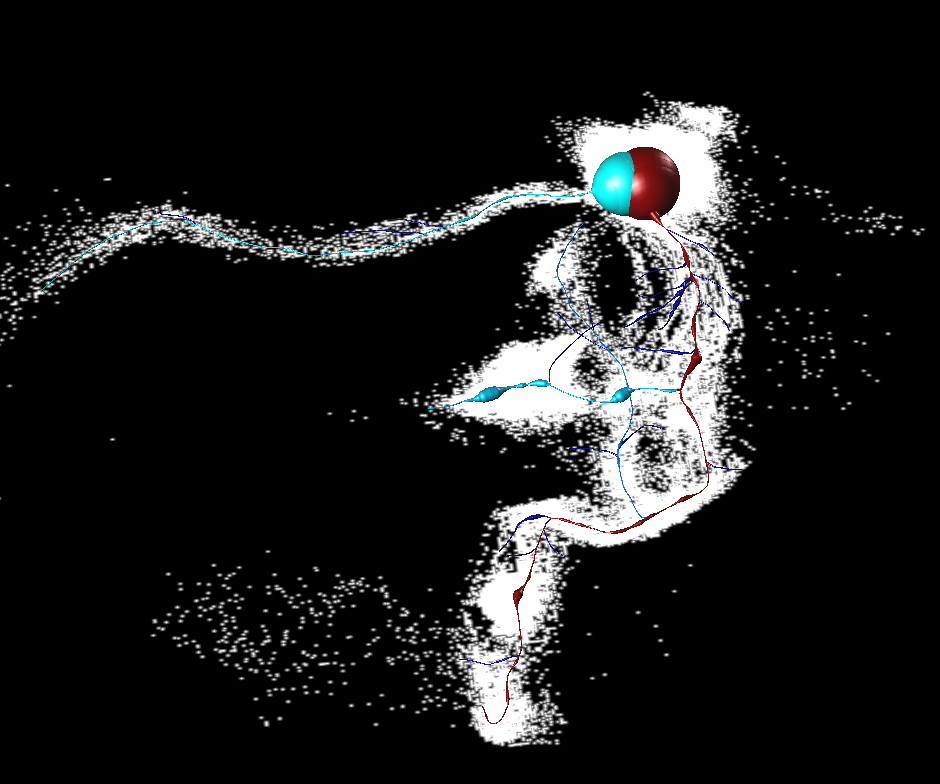} 
     & \hspace*{0.0in} &
	  \includegraphics[width=0.46\textwidth, height=6cm]{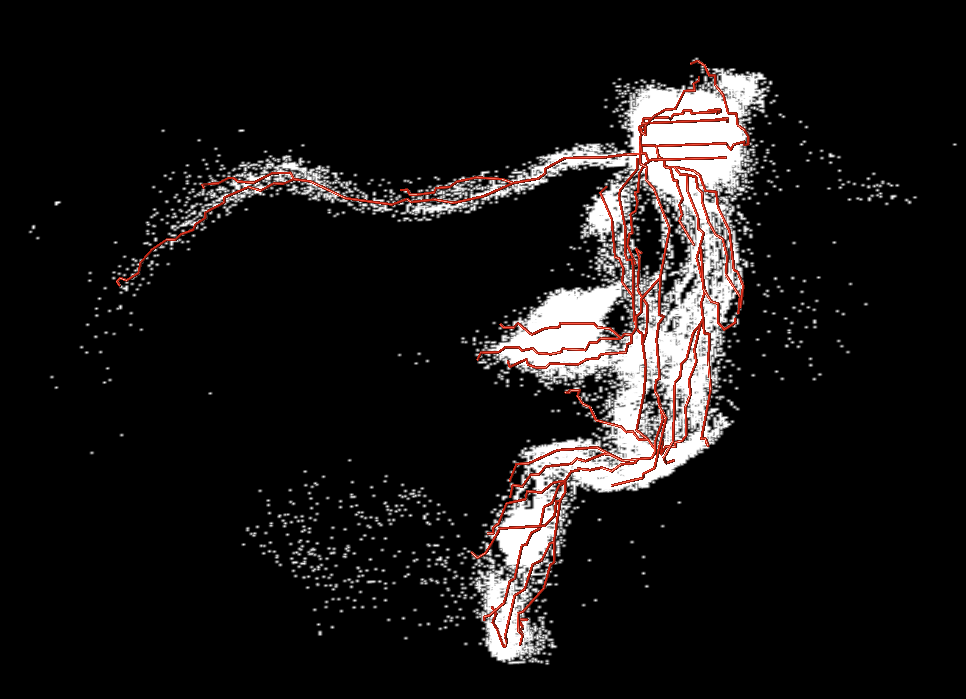} \\
(a) & & (b) \\
\includegraphics[width=0.46\textwidth, height=6cm]{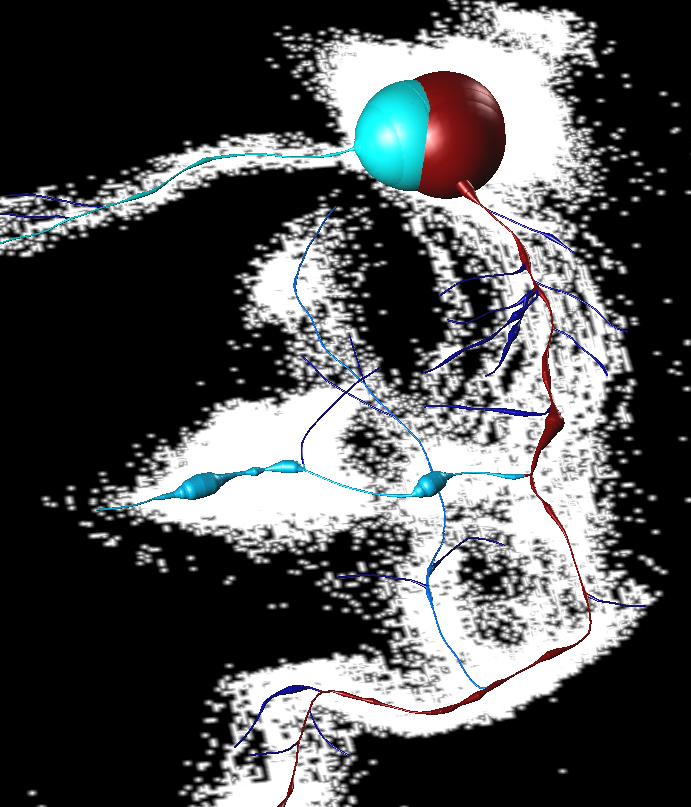} & \hspace*{0.0in} & 
\includegraphics[width=0.46\textwidth, height=6cm]{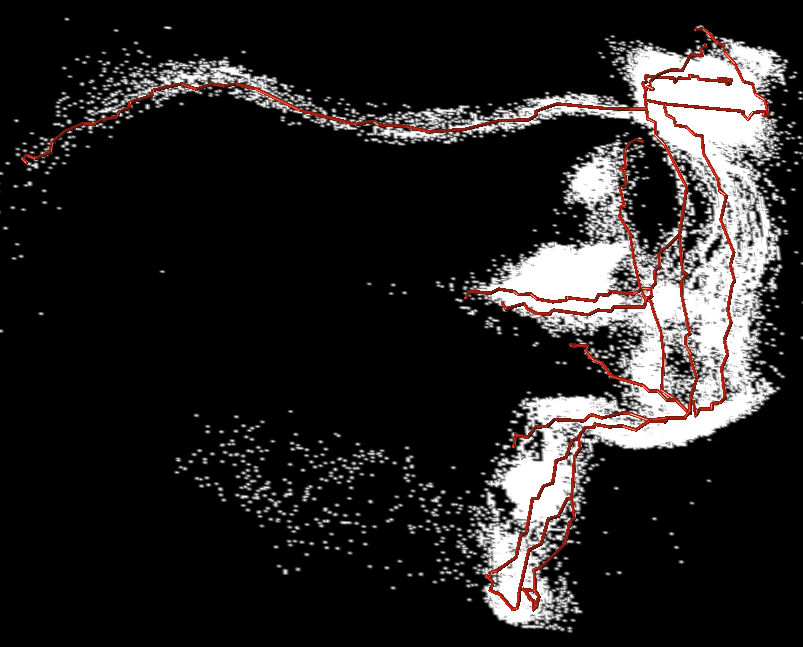} \\
(c) & (d) 
\end{tabular}
    \caption{Comparison of the result of (a) APP2, and (b) our output. We note that APP2 captures the four major branches. However, it does not capture the details: horizontal small branches are traced instead of tracing the vertical signals which are visible in the input; we show a zoomed-in view in (c). In contrast, as we increase the number of branches, our output trace those different bundles of the downwards signals better. 
    A more simplified output by our algorithm is shown in (d). }
	\label{fig:comparison}
	\end{figure}

	

	Our pipeline allows for the MBAP data to be processed in full resolution and also at a downsampled resolution.
	When processed at full resolution, the images are partitioned into tiles of 
	512*512 pixels with 5-pixels of overlapping area between adjacent tiles.
	The 1-stable manifolds are extracted from the tiles and the extraction results are merged using the stitching strategy described above. (Diffusing range 5 pixels)
	We also report the results on the downsampled image (with a resolution of 2850*3000*250) where there is no need for the divide-and-conquer approach to optimize the computing requirements. 
The results on this particular example shows that down-sampled data appears to be sufficient for summarization purposes. 
	Figure \ref{fig:LScanno} shows the results of the example LSc injection, the injection summarized tree (after simplification to show only the main trend) is shown superposed on the preprocessed input image. We have also added the  specific brain areas (blue arrows) to denote the targets of the injection. The input is a 3D point cloud generated from 2D pre-processing detection.  
    
    This summarization is annotated and aligned with BAMS database \cite{bams}, which is considered as a trust worthy standard of connectivity in rodent brains.
	Specifically in Figure \ref{fig:PMD_atlas}, each color in BAMS data represents a region connected to the injection site and our neuron summary is consistent with the BAMS data.
    
    We have applied our tree summarization framework to a collection of MBAP datasets; two examples of brains with injections in the motor cortex are shown in Figure \ref{fig:motorcortex}.
    
    In Figure \ref{fig:comparison}, we compare our output with that of APP2 (computed via software platform Vaa3D): We note that APP2 also captures the four major branches from the injection site. However, the additional branches tend to be local (semi-horizontal small branches in \ref{fig:comparison} (a) and (c)), and miss significant axonal bundles (instead, the small semi-horizontal branches cut across those bundles). We suspect that this  could be partially due to the gaps in signal. In contrast, as our algorithm uses a global view of data, our reconstruction traces the axonal bundles better; see Figure \ref{fig:comparison} (b) and (d) for two different levels of simplification. (We note that our output are guaranteed to be trees, although due to occlusion and 2D projection in the pictures it may appear that there are loops.)

	
	

\remove{
\section{Future Work}

For future work, we are exploring a better strategy for resolving loops. Currently we deal with the loops in two parts. On one hand, the persistence simplification can reliably remove small loops; on the other hand, the maximum spanning tree has a limited vision of global structures. Therefore it can happen that a loop is inappropriately cut. We have noticed that this happens in the tracer injection summarization processn when clearly defined axonal bundles are absent, indicating the need for further refinements and inclusion of biologically motivated priors. We will also explore other summarization techniques for the cloud of axonal fragments corresponding to these types of data. 
}

\bibliographystyle{abbrv}
\bibliography{reference}

\begin{thebibliography}{10}

\bibitem{Gyulassythesis}
G.~A.
\newblock Phd thesis.
\newblock {\em Univ. California Berkeley}, 2008.

\bibitem{Acciai2016}
L.~Acciai, P.~Soda, and G.~Iannello.
\newblock Automated neuron tracing methods: An updated account.
\newblock {\em Neuroinformatics}, 14(4):353--367, Oct 2016.

\bibitem{AEHW06}
P.~K. Agarwal, H.~Edelsbrunner, J.~Harer, and Y.~Wang.
\newblock Extreme elevation on a 2-manifold.
\newblock {\em Discrete and Computational Geometry (DCG)}, 36(4):553--572,
  2006.

\bibitem{Al-Kofahi:2002}
K.~A. Al-Kofahi, S.~Lasek, D.~H. Szarowski, C.~J. Pace, G.~Nagy, J.~N. Turner,
  and B.~Roysam.
\newblock Rapid automated three-dimensional tracing of neurons from confocal
  image stacks.
\newblock {\em Trans. Info. Tech. Biomed.}, 6(2):171--187, June 2002.

\bibitem{arenkiel2014neural}
B.~Arenkiel.
\newblock {\em Neural Tracing Methods: Tracing Neurons and Their Connections}.
\newblock Neuromethods. Springer New York, 2014.

\bibitem{Bas2011}
E.~Bas and D.~Erdogmus.
\newblock Principal curves as skeletons of tubular objects.
\newblock {\em Neuroinformatics}, 9(2):181--191, Sep 2011.

\bibitem{autofluoro}
W.~Baschong, R.~Suetterlin, and R.~H. Laeng.
\newblock Control of autofluorescence of archival formaldehyde-fixed,
  paraffin-embedded tissue in confocal laser scanning microscopy (clsm).
\newblock {\em Journal of Histochemistry \& Cytochemistry}, 49(12):1565--1571,
  2001.
\newblock PMID: 11724904.

\bibitem{Basu2014}
S.~Basu, W.~T. Ooi, and D.~Racoceanu.
\newblock Improved marked point process priors for single neurite tracing.
\newblock pages 1--4, 06 2014.

\bibitem{BAUER2017}
U.~Bauer, M.~Kerber, J.~Reininghaus, and H.~Wagner.
\newblock Phat – persistent homology algorithms toolbox.
\newblock {\em Journal of Symbolic Computation}, 78(Supplement C):76 -- 90,
  2017.
\newblock Algorithms and Software for Computational Topology.

\bibitem{Boykov2001}
Y.~Boykov, O.~Veksler, and R.~Zabih.
\newblock Fast approximate energy minimization via graph cuts.
\newblock {\em IEEE Transactions on Pattern Analysis and Machine Intelligence},
  23(11):1222--1239, Nov 2001.

\bibitem{Carlsson:2004}
G.~Carlsson, A.~Zomorodian, A.~Collins, and L.~Guibas.
\newblock Persistence barcodes for shapes.
\newblock In {\em Proceedings of the 2004 Eurographics/ACM SIGGRAPH Symposium
  on Geometry Processing}, SGP '04, pages 124--135, New York, NY, USA, 2004.
  ACM.

\bibitem{smarttracing}
H.~Chen, H.~Xiao, T.~Liu, and H.~Peng.
\newblock Smarttracing: self-learning-based neuron reconstruction.
\newblock {\em Brain Informatics}, 2(3):135--144, 2015.

\bibitem{Choromanska2012}
A.~Choromanska, S.-F. Chang, and R.~Yuste.
\newblock Automatic reconstruction of neural morphologies with multi-scale
  tracking.
\newblock {\em Frontiers in Neural Circuits}, 6:25, 2012.

\bibitem{Chothani2011}
P.~Chothani, V.~Mehta, and A.~Stepanyants.
\newblock Automated tracing of neurites from light microscopy stacks of images.
\newblock {\em Neuroinformatics}, 9(2):263--278, Sep 2011.

\bibitem{chung}
M.~K. Chung, P.~Bubenik, and P.~T. Kim.
\newblock Persistence diagrams of cortical surface data.
\newblock In J.~L. Prince, D.~L. Pham, and K.~J. Myers, editors, {\em
  Information Processing in Medical Imaging}, pages 386--397, Berlin,
  Heidelberg, 2009. Springer Berlin Heidelberg.

\bibitem{DRS15}
O.~Delgado-Friedrichs, V.~Robins, and A.~Sheppard.
\newblock Skeletonization and partitioning of digital images using discrete
  morse theory.
\newblock {\em IEEE Trans. Pattern Anal. Machine Intelligence}, 37(3):654--666,
  March 2015.

\bibitem{DWW17}
T.~K. dey, J.~Wang, and Y.~Wang.
\newblock Improved road network reconstruction using discrete morse theory.
\newblock In {\em Proceedings of the 25th ACM SIGSPATIAL International
  Conference on Advances in Geographic Information Systems (ACM SIGSPATIAL
  GIS)}, page~58, 2017.

\bibitem{DWW18}
T.~K. Dey, J.~Wang, and Y.~Wang.
\newblock Graph reconstruction by discrete {Morse} theory.
\newblock In {\em Sympos. Comput. Geometry (SoCG), to appear}, 2018.

\bibitem{DONOHUE201194}
D.~E. Donohue and G.~A. Ascoli.
\newblock Automated reconstruction of neuronal morphology: An overview.
\newblock {\em Brain Research Reviews}, 67(1):94 -- 102, 2011.

\bibitem{Edelsbrunner2002}
Edelsbrunner, Letscher, and Zomorodian.
\newblock Topological persistence and simplification.
\newblock {\em Discrete {\&} Computational Geometry}, 28(4):511--533, Nov 2002.

\bibitem{Edelsbrunner_persistenthomology}
H.~Edelsbrunner and J.~Harer.
\newblock Persistent homology -- a survey.

\bibitem{persistsurvey}
M.~{Ferri}.
\newblock {Persistent topology for natural data analysis - A survey}.
\newblock {\em ArXiv e-prints}, June 2017.

\bibitem{Frangi1998}
A.~F. Frangi, W.~J. Niessen, K.~L. Vincken, and M.~A. Viergever.
\newblock {\em Multiscale vessel enhancement filtering}, pages 130--137.
\newblock Springer Berlin Heidelberg, Berlin, Heidelberg, 1998.

\bibitem{activelearning2014}
R.~Gala, J.~Chapeton, J.~Jitesh, C.~Bhavsar, and A.~Stepanyants.
\newblock Active learning of neuron morphology for accurate automated tracing
  of neurites.
\newblock {\em Frontiers in Neuroanatomy}, 8:37, 2014.

\bibitem{diadem_metric}
T.~A. Gillette, K.~M. Brown, and G.~A. Ascoli.
\newblock The diadem metric: Comparing multiple reconstructions of the same
  neuron.
\newblock {\em Neuroinformatics}, 9(2):233, Apr 2011.

\bibitem{GDN07}
A.~Gyulassy, M.~Duchaineau, V.~Natarajan, V.~Pascucci, E.~Bringa,
  A.~Higginbotham, and B.~Hamann.
\newblock Topologically clean distance fields.
\newblock {\em IEEE Trans. Visualization Computer Graphics}, 13(6):1432--1439,
  Nov 2007.

\bibitem{Lamar}
J.~Lamar-Le{\'o}n, E.~B. Garc{\'i}a-Reyes, and R.~Gonzalez-Diaz.
\newblock Human gait identification using persistent homology.
\newblock In L.~Alvarez, M.~Mejail, L.~Gomez, and J.~Jacobo, editors, {\em
  Progress in Pattern Recognition, Image Analysis, Computer Vision, and
  Applications}, pages 244--251, Berlin, Heidelberg, 2012. Springer Berlin
  Heidelberg.

\bibitem{Lee2008}
P.-C. Lee, Y.-T. Ching, H.~M. Chang, and A.-S. Chiang.
\newblock A semi-automatic method for neuron centerline extraction in confocal
  microscopic image stack.
\newblock In {\em 2008 5th IEEE International Symposium on Biomedical Imaging:
  From Nano to Macro}, pages 959--962, May 2008.

\bibitem{lee2012}
P.-C. Lee, C.-C. Chuang, A.-S. Chiang, and Y.-T. Ching.
\newblock High-throughput computer method for 3d neuronal structure
  reconstruction from the image stack of the drosophila brain and its
  applications.
\newblock {\em PLOS Computational Biology}, 8(9):1--12, 09 2012.

\bibitem{Liu2016ApplyingTP}
J.-Y. Liu, S.-K. Jeng, and Y.-H. Yang.
\newblock Applying topological persistence in convolutional neural network for
  music audio signals.
\newblock {\em CoRR}, abs/1608.07373, 2016.

\bibitem{neuronsurvey}
E.~Meijering.
\newblock Neuron tracing in perspective.
\newblock {\em Cytometry Part A}, 77A(7):693--704, 2010.

\bibitem{Morse_Theory}
J.~Milnor.
\newblock {\em Morse Theory}.
\newblock Princeton Univ. Press, New Jersey, 1963.

\bibitem{PMDmouse}
Brain architecture project.
\newblock http://mouse.brainarchitecture.org/homepage/.

\bibitem{bams}
Bams dataset.
\newblock https://bams1.org/.

\bibitem{Myatt2012}
D.~Myatt, T.~Hadlington, G.~Ascoli, and S.~Nasuto.
\newblock Neuromantic – from semi-manual to semi-automatic reconstruction of
  neuron morphology.
\newblock {\em Frontiers in Neuroinformatics}, 6:4, 2012.

\bibitem{oh2014mesoscale}
S.~W. Oh, J.~A. Harris, L.~Ng, B.~Winslow, N.~Cain, S.~Mihalas, Q.~Wang,
  C.~Lau, L.~Kuan, A.~M. Henry, et~al.
\newblock A mesoscale connectome of the mouse brain.
\newblock {\em Nature}, 508(7495):207--214, 2014.

\bibitem{pengapp}
H.~Peng, F.~Long, and G.~Myers.
\newblock Automatic 3d neuron tracing using all-path pruning.
\newblock {\em Bioinformatics}, 27(13):i239, 2011.

\bibitem{platt}
D.~Platt, S.~Basu, P.~Zalloua, and L.~Parida.
\newblock Characterizing redescriptions using persistent homology to isolate
  genetic pathways contributing to pathogenesis.
\newblock 10, 01 2016.

\bibitem{diademchallenge}
Diadem challenge.
\newblock http://diademchallenge.org.

\bibitem{DiMorSC}
Dimorsc.
\newblock http://github.com/SuyiWang/DiMorSC.

\bibitem{Vaa3D}
Vaa3d.
\newblock http://vaa3d.org.

\bibitem{RWS11}
V.~Robins, P.~J. Wood, and A.~P. Sheppard.
\newblock Theory and algorithms for constructing discrete morse complexes from
  grayscale digital images.
\newblock {\em IEEE Trans. Pattern Anal. Machine Intelligence},
  33(8):1646--1658, Aug 2011.

\bibitem{SCHMITT20041283}
S.~Schmitt, J.~F. Evers, C.~Duch, M.~Scholz, and K.~Obermayer.
\newblock New methods for the computer-assisted 3-d reconstruction of neurons
  from confocal image stacks.
\newblock {\em NeuroImage}, 23(4):1283 -- 1298, 2004.

\bibitem{sironi2015}
A.~Sironi, V.~Lepetit, and P.~Fua.
\newblock Projection onto the manifold of elongated structures for accurate
  extraction.
\newblock In {\em 2015 IEEE International Conference on Computer Vision
  (ICCV)}, pages 316--324, Dec 2015.

\bibitem{2011MNRAS}
T.~{Sousbie}.
\newblock {The persistent cosmic web and its filamentary structure - I. Theory
  and implementation}.
\newblock 414:350--383, June 2011.

\bibitem{Srinivasan2007}
R.~Srinivasan, X.~Zhou, E.~Miller, J.~Lu, J.~Litchman, and S.~T.~C. Wong.
\newblock Automated axon tracking of 3d confocal laser scanning microscopy
  images using guided probabilistic region merging.
\newblock {\em Neuroinformatics}, 5(3):189--203, Sep 2007.

\bibitem{sui2014}
D.~Sui, K.~Wang, J.~Chae, Y.~Zhang, and H.~Zhang.
\newblock A pipeline for neuron reconstruction based on spatial sliding volume
  filter seeding.
\newblock 2014:386974, 07 2014.

\bibitem{Turetken2011}
E.~T{\"u}retken, G.~Gonz{\'a}lez, C.~Blum, and P.~Fua.
\newblock Automated reconstruction of dendritic and axonal trees by global
  optimization with geometric priors.
\newblock {\em Neuroinformatics}, 9(2):279--302, Sep 2011.

\bibitem{turetken2013}
E.~Türetken, F.~Benmansour, B.~Andres, H.~Pfister, and P.~Fua.
\newblock Reconstructing loopy curvilinear structures using integer
  programming.
\newblock In {\em 2013 IEEE Conference on Computer Vision and Pattern
  Recognition}, pages 1822--1829, June 2013.

\bibitem{VASILKOSKI2009197}
Z.~Vasilkoski and A.~Stepanyants.
\newblock Detection of the optimal neuron traces in confocal microscopy images.
\newblock {\em Journal of Neuroscience Methods}, 178(1):197 -- 204, 2009.

\bibitem{Suyithesis}
S.~Wang.
\newblock {\em Analyzing data with 1D non-linear shapes using topological
  methods}.
\newblock PhD thesis, The Ohio State University, Computer Science and
  Engineering Department, 2018.

\bibitem{Wang2015}
S.~Wang, Y.~Wang, and Y.~Li.
\newblock Efficient map reconstruction and augmentation via topological
  methods.
\newblock In {\em Proceedings of the 23rd SIGSPATIAL International Conference
  on Advances in Geographic Information Systems}, SIGSPATIAL '15, pages
  25:1--25:10, New York, NY, USA, 2015. ACM.

\bibitem{Wang2011}
Y.~Wang, A.~Narayanaswamy, C.-L. Tsai, and B.~Roysam.
\newblock A broadly applicable 3-d neuron tracing method based on open-curve
  snake.
\newblock {\em Neuroinformatics}, 9(2):193--217, 2011.

\bibitem{app2}
H.~Xiao and H.~Peng.
\newblock App2: automatic tracing of 3d neuron morphology based on hierarchical
  pruning of a gray-weighted image distance-tree.
\newblock {\em Bioinformatics}, 29(11):1448, 2013.

\bibitem{Yang2013}
J.~Yang, P.~T. Gonzalez-Bellido, and H.~Peng.
\newblock A distance-field based automatic neuron tracing method.
\newblock {\em BMC Bioinformatics}, 14(1):93, Mar 2013.

\bibitem{Yuan2009}
X.~Yuan, J.~T. Trachtenberg, S.~M. Potter, and B.~Roysam.
\newblock Mdl constrained 3-d grayscale skeletonization algorithm for automated
  extraction of dendrites and spines from fluorescence confocal images.
\newblock {\em Neuroinformatics}, 7(4):213, 2009.

\bibitem{ZHANG2007149}
Y.~Zhang, X.~Zhou, A.~Degterev, M.~Lipinski, D.~Adjeroh, J.~Yuan, and S.~T.
  Wong.
\newblock A novel tracing algorithm for high throughput imaging: Screening of
  neuron-based assays.
\newblock {\em Journal of Neuroscience Methods}, 160(1):149 -- 162, 2007.

\bibitem{Zhao2011}
T.~Zhao, J.~Xie, F.~Amat, N.~Clack, P.~Ahammad, H.~Peng, F.~Long, and E.~Myers.
\newblock Automated reconstruction of neuronal morphology based on local
  geometrical and global structural models.
\newblock {\em Neuroinformatics}, 9(2):247--261, Sep 2011.

\bibitem{Zhou1999}
Y.~Zhou and A.~W. Toga.
\newblock Efficient skeletonization of volumetric objects.
\newblock {\em IEEE Transactions on Visualization and Computer Graphics},
  5(3):196--209, Jul 1999.

\bibitem{Zhou2016}
Z.~Zhou, X.~Liu, B.~Long, and H.~Peng.
\newblock Tremap: Automatic 3d neuron reconstruction based on tracing, reverse
  mapping and assembling of 2d projections.
\newblock {\em Neuroinformatics}, 14(1):41--50, Jan 2016.

\bibitem{Zhou2015}
Z.~Zhou, S.~A. Sorensen, and H.~Peng.
\newblock Neuron crawler: An automatic tracing algorithm for very large neuron
  images.
\newblock In {\em 2015 IEEE 12th International Symposium on Biomedical Imaging
  (ISBI)}, pages 870--874, April 2015.

\bibitem{zomorodian_2005}
A.~J. Zomorodian.
\newblock {\em Topology for Computing}.
\newblock Cambridge Monographs on Applied and Computational Mathematics.
  Cambridge University Press, 2005.

\end{thebibliography}

\end{document}